\documentclass[reprint,showpacs,amsmath,amssymb,prc,superscriptaddress,nofootinbib]{revtex4-1}

\usepackage{graphicx}

\usepackage{dcolumn}
\usepackage{bm}
\usepackage{times}

\newcolumntype{.}{D{.}{.}{1}}
\newcolumntype{X}{D{X}{X}{1}}

\begin{document}

\newcommand{\reaction}[6]{\nuc{#1}{#2}(#3,#4)\/\nuc{#5}{#6}}
\newcommand{\nuc}[2]{\ensuremath{^{#1}}#2}
\newcommand{\Ne}[0]{\ensuremath{^{22}}Ne }
\newcommand{\Alpp}[0]{\ensuremath{^{27}}Al\ensuremath{+}p }
\newcommand{\Nepp}[0]{\ensuremath{^{22}}Ne\ensuremath{+}p }
\newcommand{\Nepa}[0]{\ensuremath{^{22}}Ne\ensuremath{+ \alpha} }
\newcommand{\Erlab}[1]{\ensuremath{E_{r}^{\text{lab}} = #1}~keV}
\newcommand{\Erlabb}[1]{\ensuremath{\mathbf{E_{r}^{\textbf{lab}} = #1}}~\textbf{keV}}
\newcommand{\Ex}[1]{\ensuremath{E_{x} = #1}~keV}
\newcommand{\Exb}[1]{\ensuremath{\mathbf{E_{x} = #1}}~\textbf{keV}}
\newcommand{\ag}[0]{\reaction{22}{Ne}{$\alpha$}{$\gamma$}{26}{Mg} }
\newcommand{\an}[0]{\reaction{22}{Ne}{$\alpha$}{n}{25}{Mg} }
\newcommand{\Jpi}[2]{\ensuremath{J^{\pi} = #1^{#2}} }
\newcommand{\Jpib}[2]{\ensuremath{\mathbf{J^{\pi} = #1^{#2}}} }
\def\mggg{\ensuremath{^{26}}Mg\ensuremath{(\Vec{\gamma},\gamma)^{26}}Mg }
\def\gray{\ensuremath{\gamma}-ray}
\def\sun{\odot}

\title{Reaction rates for the s-process neutron source \Nepa}

\author{R.~Longland}
\affiliation{Departament de F\'{i}sica i Enginyeria Nuclear, EUETIB,
  Universitat Polit\`{e}cnica de Catalunya, 08036 Barcelona, Spain}
\affiliation{Department of Physics and Astronomy, University of North
  Carolina at Chapel Hill, Chapel Hill, NC 27599, USA}
\affiliation{Triangle Universities Nuclear Laboratory, Durham, NC
  27708, USA}

\author{C.~Iliadis} 
\affiliation{Department of Physics and Astronomy, University of North
  Carolina at Chapel Hill, Chapel Hill, NC 27599, USA}
\affiliation{Triangle Universities Nuclear Laboratory, Durham, NC
  27708, USA}

\author{A.~I.~Karakas}
\affiliation{Research School of Astronomy \& Astrophysics, Mount
  Stromlo Observatory, Weston Creek ACT 2611, Australia}


\begin{abstract}
  The \an reaction is an important source of neutrons for the
  s-process. In massive stars responsible for the weak component of
  the s-process, \an is the dominant source of neutrons, both during
  core helium burning and in shell carbon burning. For the main
  s-process component produced in Asymptotic Giant Branch (AGB)
  stars, the \reaction{13}{C}{$\alpha$}{n}{16}{O} reaction is the
  dominant source of neutrons operating during the interpulse period,
  with the \Nepa source affecting mainly the s-process branchings
  during a thermal pulse. Rate uncertainties in the competing \an and
  \ag reactions result in large variations of s-process
  nucleosynthesis. Here, we present up-to-date and statistically
  rigorous \Nepa reaction rates using recent experimental results and
  Monte Carlo sampling. Our new rates are used in post-processing
  nucleosynthesis calculations both for massive stars and AGB stars. We
  demonstrate that the nucleosynthesis uncertainties arising from the
  new rates are dramatically reduced in comparison to previously
  published results, but several ambiguities in the present data must
  still be addressed. Recommendations for further study to resolve
  these issues are provided.
\end{abstract}


\maketitle


\section{Introduction}
\label{sec:intro}

The s-process is responsible for creating about half of the elements
heavier than iron that are observed in the solar system\
\citep{SNE08}. This process involves the slow capture of neutrons
(slower than the average $\beta$-decay rate of unstable nuclei) onto
seed material, hence nucleosynthesis follows the nuclear valley of
stability. By considering the solar system abundances of s-only nuclei
(that is, nuclei that can only be produced in the s-process) it can be
shown that there are two key components of the s-process: the ``main''
component and the ``weak'' component\ \citep{KAP11}. The main
component produces s-nuclei with masses of $A>90$, while the weak
component enriches the s-nuclei abundances at $A \lesssim 90$.

The main component of the s-process arises from neutron captures
during He-burning in $M \leq 4 M_{\odot}$ Asymptotic Giant Branch
(AGB) stars (a detailed discussion of nuclear burning in AGB stars can
be found in Refs.~\cite{AGBBook} and \cite{HER05}). In low mass (0.8 to 4\
$M_{\sun}$) AGB stars of solar metalicity, most neutrons are released
through the $^{13}$C($\alpha$,n)$^{16}$O reaction during the
inter-pulse period, while the \an reaction produces an additional
burst of neutrons during thermal pulses. This burst of neutrons
affects mainly the branchings in the s-process path. In
intermediate-mass AGB stars ($M>4M_{\odot}$), where the temperatures
are expected to be higher, the \an reaction is thought to be the main
source of neutrons and could explain the enhancement of rubidium seen
in some metal poor AGB stars
\citep{PIG05,GAR06,LUG08,GAR09,KAR12}. 
In addition to s-process elements, the \an and \ag rates influence the
relative production of \nuc{25}{Mg} and \nuc{26}{Mg}, whose abundance
ratios can be measured to high precision in circumstellar
(``presolar'') dust grains. Magnesium is also one of the few elements
for which the isotopic ratios (\nuc{25}{Mg}/\nuc{24}{Mg} and
\nuc{26}{Mg}/\nuc{24}{Mg} can be derived from stellar spectra (for
example, Refs.~\cite{YON03a,YON03b}). However, \textcite{KAR06} showed
that with their estimated \an and \ag reaction rate uncertainties, the
relative abundances of \nuc{25}{Mg} and \nuc{26}{Mg} predicted by
their stellar models can vary by up to 60\%.

The weak component of the s-process arises from nuclear burning in
massive stars. The core temperature in these stars (typically with $M
\gtrsim 11M_{\odot}$) becomes high enough during He-burning for the
\an reaction to produce a high flux of neutrons shortly before the
helium fuel is exhausted. Any remaining \nuc{22}{Ne} releases a second
flux of neutrons during convective carbon shell burning. The s-process
yield in these stars is therefore sensitive to the temperature at
which the \an reaction starts to produce an appreciable flux of
neutrons. \textcite{THE07} showed that the s-process during the core
He-burning stage in massive stars depends strongly on both the \Nepa
and the \reaction{16}{O}{n}{$\gamma$}{17}{O} reaction rates. They also
found that not only are the overall uncertainties in the rates
important, but also the temperature dependence of the
rates. 

The \Nepa reactions also affect nucleosynthesis in other
astrophysical environments. During type II supernova explosions, two
$\gamma$-ray emitting radionuclides, \nuc{26}{Al} and \nuc{60}{Fe} are
ejected, 
and their abundance ratio provides a sensitive constraint on stellar
models \cite[][and references therein]{WOO07}. The species
\nuc{60}{Fe} is mainly produced in massive stars by neutron captures
during convective shell carbon burning\ \citep[e.g.,][]{PIG10}. Its
abundance, therefore, depends strongly on the \Nepa rates. The \Nepa
rates also play a role in type Ia supernovae. Throughout the
``simmering'' stage, roughly 1000 years prior to the explosion,
\textcite{PIR09} suggested that neutrons released by the \an reaction
affect the carbon abundance, thus altering the amount of \nuc{56}{Ni}
produced (i.e., the peak luminosity) in the explosion. \textcite{TIM03}
also found that during the explosion, neutronisation by the \an
reaction affects the electron mole fraction, $Y_e$, thus influencing
the nature of the explosion.

In this work we will evaluate new reaction rates for
\mbox{\Nepa}. Compared to previous results\ \citep{ANG99,JAE01,KAR06}
our new rates are significantly improved because (i) we incorporate
all the recently obtained data on resonance fluorescence absorption,
$\alpha$-particle transfer etc., and (ii) we employ a sophisticated
(Monte Carlo) method to estimate the rates and associated
uncertainties. 
We have recently presented new \Nepa rates in Ref.\ \cite{ILI10a}, but did
not give a detailed account of their calculation. Since the latter
results were published, we found, and could account for, a number of
inconsistencies in data previously reported in the literature. In
addition, new data from Ref.\ \cite{DEB10} became available, which have been
included in the present work. Thus the rates presented here supersede
our earlier results\ \citep{ILI10a}.

The paper will be organised as follows: in Sec.\ \ref{sec:formalism} a
detailed discussion of the Monte Carlo method used to calculate
reaction rates is discussed. This method is described in detail
elsewhere\ \citep{LON10} but will be summarised to show its
applicability to the specific cases of the \Nepa reactions. The \Nepa
rate calculations and comparisons with the literature will
be presented in Sec.\ \ref{sec:rates}. The reaction rates will then be
used to present new nucleosynthesis yields along with their
uncertainties in Sec.~\ref{sec:results}. Conclusions will be
presented in Sec.\ \ref{sec:conclusions}.

\section{Reaction Rate Formalism}
\label{sec:formalism}
\subsection{Thermonuclear Reaction Rates}
\label{sec:rates-formalism}
The reaction rate per particle pair in a plasma of temperature, $T$,
is given by
\begin{equation}
  \label{eq:rates-reacrate}
  \langle\sigma v\rangle = \sqrt{\frac{8}{\pi \mu}}
  \frac{1}{(kT)^{3/2}}\int_{0}^{\infty}E\sigma(E)e^{-E/kT} dE
\end{equation}
where $\mu$ is the reduced mass of the reacting particles,
$\mu=M_{0}M_{1}/(M_0+M_1)$; $M_i$ denotes the masses of the
particles; $k$ is the Boltzmann constant; $E$ is the centre-of-mass
energy of the reacting particles; and $\sigma(E)$ is the reaction
cross section at energy, $E$. 

The strategy for determining reaction rates from Eq.\
(\ref{eq:rates-reacrate}) depends on the nature of the cross section.
In many cases the cross section can be separated into non-resonant and
resonant parts. Reactions such as \Nepa proceed through the compound
nucleus \nuc{26}{Mg} at relatively high excitation energy ($Q_{\alpha
  \gamma}=10614.787 (33)$\ keV\ \citep{AME03}) and are frequently
dominated by resonant capture. The non-resonant part of the cross
section will, therefore, be neglected in the following discussion. The
reader is referred to Refs.\ \cite{ILIBook} and \cite{LON10} for more
details.

The resonant part of the cross-section can be represented in one of
two ways: (i) by narrow resonances, whose partial widths can be
assumed to be approximately constant over the resonance width
(``narrow resonances''), and (ii) by wide resonances, for which the
resonant cross section must be integrated numerically to account for
the energy dependence of the partial widths involved. The reaction
rates per particle pair for single, isolated narrow and wide
resonances, respectively, are given by
\begin{equation}
  \label{eq:rates-narrowrate}
  \langle\sigma v\rangle = \left(\frac{2\pi}{\mu
      kT}\right)^{3/2}\hbar^2 \omega\gamma e^{-E_r/kT}
\end{equation}
and
\begin{equation}
  \label{eq:rates-rr-resonance}
  \langle\sigma v\rangle = 
  \frac{\sqrt{2\pi}\hbar^2}{(\mu kT)^{3/2}}\omega \int_{0}^{\infty}
  \frac{\Gamma_a(E)\Gamma_b(E)}{(E-E_r)^2 + \Gamma(E)^2/4} e^{-E/kT} dE
\end{equation}
where the resonance strength, $\omega \gamma$ is defined by
\begin{equation}
  \label{eq:rates-ResStrength}
  \omega \gamma = \omega \Gamma_a \Gamma_b/\Gamma,
\end{equation}
$E_r$ is the resonance energy; $\Gamma_a(E)$, $\Gamma_b(E)$, and
$\Gamma(E)$ are the energy-dependent entrance channel (particle)
partial width, exit channel partial width, and total width,
respectively; and $\omega$, the statistical spin factor, is defined
by $\omega = (2J+1)/(2J_0+1)(2J_1+1)$, where $J$ and $J_i$ are the
resonance and particle spins, respectively. The particle partial
width, $\Gamma_c$, can be written as the product of an
energy-independent reduced width, $\gamma_c^2$, and an
energy-dependent penetration factor, $P_c(E)$, as
\begin{equation}
  \label{eq:rates-GammaDefinition}
  \Gamma_{c} = 2 P_c(E) \gamma_c^2.
\end{equation}
For the present case of $^{22}$Ne$+\alpha$, the entrance channel
($\alpha$-particle) reduced width, $\gamma^2_{\alpha}$, is related to
the $\alpha$-particle spectroscopic factor, $S_{\alpha}$, by 
\begin{align}
  \label{eq:S-reduced-width-relation1}
  \gamma^2_{\alpha} =& \frac{\hbar^2}{\mu a^2} \theta^2_{\alpha} \\
  \label{eq:S-reduced-width-relation2}
  & = \frac{\hbar^2}{2 \mu a} S_{\alpha}\phi^2(a)
\end{align}
where $\phi(a)$ is the single-particle radial wave function at the
channel radius, $a$~\citep[see, for example,][]{ILI97,BEL07}. The
constant $\hbar^2/(\mu a^2)$ is the Wigner Limit (in the notation of
Lane and Thomas~\cite{LAN58}). It can be regarded as an upper limit,
according to the sum rules in the dispersion theory of nuclear
reactions, i.e., $\theta^2_{\alpha} \leq 1$. Note that it is
frequently assumed that $\theta^2=S$, which must be regarded as a
crude approximation only. For example, in the case of $^{17}$O levels
it was shown in Ref.~\cite{KEE03b} that $S_{\alpha}$ exceeds
$\theta^2_{\alpha}$ by at most a factor of 2. We will return to these
issues in Sec. \ref{sec:spectro-factors}.

The above relationships are useful since they allow for an estimation
of the important $\alpha$-particle partial widths from spectroscopic
factors obtained in $\alpha$-particle transfer reactions, as will be
discussed later. It is important to note that the value of
$S_{\alpha}$ depends on the parameters of the nuclear potentials
assumed in the transfer data analysis. Similarly, the value of
$\gamma^2_{\alpha}$ depends on the channel radius. However, if,
throughout the analysis, consistent values of these parameters (such
as $a$) are used, their impact on the value of $\Gamma_{\alpha}$ will
be strongly reduced.

\subsection{Monte Carlo Reaction Rates}
\label{sec:rates-montecarlo}

The equations outlined in Sec.\ \ref{sec:rates-formalism} provide the
tools for calculating thermonuclear reaction rates given available
estimates for the cross section parameters ($E_r$, $\omega\gamma$,
etc.). A problem arises, however, when statistically rigorous
uncertainties of the reaction rates are desired.  What is usually
presented in the literature are recommended rates, together with upper
and lower ``limits'', but the reported values are not derived from a
suitable probability density function. Therefore, the reported values
have no rigorous statistical meaning.  An attempt to construct a
method for analytical uncertainty propagation of reaction rates was
made by \textcite{THO99}. However, their method is applicable only in
special cases, when the uncertainties in resonance parameters are
relatively small. \textcite{THO99} were also not able to treat the
uncertainty propagation for reaction rates that need to be integrated
or for rates that include upper limits on some parameters. For these
reasons, a Monte Carlo method is used in the present study to
calculate statistically meaningful reaction rates.

The general strategy of Monte Carlo uncertainty\footnote{Throughout
  this work, care is taken to refer to the terms \textit{uncertainty}
  and \textit{error} correctly. The term \textit{error} refers to a
  quantity that is believed to be incorrect, whereas
  \textit{uncertainty} refers to the spread estimate of a parameter.}
propagation is as follows: (i) randomly sample from the probability
density distribution of each input parameter; (ii) calculate the
reaction rates for each randomly sampled parameter set on a grid of
temperatures (using the same set at each temperature); (iii) repeat
steps (i)-(ii) many times (on the order of 5000).  Steps (i)--(iii)
will result in a distribution at each temperature grid point that can
be interpreted as the probability density function of the reaction
rate.  Extraction of uncertainties from this distribution will be
discussed later. While input parameter sampling is being performed,
care must be taken to consider correlations in parameters. For
example, particle partial widths depend on the penetration factor,
which is an energy dependent quantity.  The individual energy samples
must, therefore, be propagated consistently through resonance energy
and partial width estimation in order to fully account for the
correlation of these quantities. The code \texttt{RatesMC}\
\citep{LON10} was used to perform the Monte Carlo sampling and to
analyse the probability densities of the total reaction rates.

In order to apply Monte Carlo sampling to calculate reaction rate
uncertainties, sampling distributions must be chosen for each input
parameter. Once a reaction rate (output) distribution has been
computed, an appropriate mathematical description must be found to
present the result in a convenient manner. Statistical distributions
important for reaction rate calculations are described in detail in
Refs.\ \cite{LON10} and \cite{LON10T}, and are summarised briefly
below.

Uncertainties of resonance energies are determined by the \textit{sum}
of different contributions. In this case, the central limit theorem of
statistics predicts that resonance energies are Gaussian distributed.
Note that there is a finite probability of calculating a negative
resonance energy and that this choice of probability density naturally
accounts for the inclusion of sub-threshold resonances in the above
formalism.
A resonance strength or a partial width, on the other hand, is
experimentally derived from the \textit{product} of measured input
quantities (e.g., count rates, stopping powers, detection
efficiencies, etc.). In such a case the central limit theorem predicts
that resonance strengths or partial widths are lognormally
distributed.

The lognormal probability density for a resonance strength or a
partial width is given by
\begin{equation}
  \label{eq:rates-lognormal}
  f(x) = \frac{1}{\sigma \sqrt{2 \pi}} \frac{1}{x} e^{-(\ln x -
    \mu)^2/(2 \sigma^2)}
\end{equation}
with the lognormal parameters $\mu$ and $\sigma$ representing the mean
and standard deviation of $\ln{x}$. These quantities are related to
the expectation value, $E[x]$, and variance, $V[x]$, by
\begin{equation}
  \label{eq:rates-lognormpars}
  \mu = \ln(E[x]) - \frac{1}{2}\ln\left(1+\frac{V[x]}{E[x]^2}\right),
  \quad \sigma = \sqrt{\ln\left(1+\frac{V[x]}{E[x]^2}\right)}
\end{equation}
The quantities, $\ln(E[x])$ and $\sqrt V[x]$ can be associated with
the central value and uncertainty, respectively, that are commonly
reported. Note that a lognormal distribution is only defined for
positive values of $x$. This feature is crucial because it removes the
finite probability of sampling unphysical, negative values when
Gaussian uncertainties are used. This is especially true for partial
width measurements, which frequently have uncertainties in the 20-50\%
range. Note, also, that a 50\% Gaussian uncertainty results in a 3\%
probability of the partial width having a value below zero.

The important problem of estimating reaction rates when only upper
limits of resonance strengths or partial widths are available will now
be discussed. The standard practise in nuclear
astrophysics~\citep[see, for example,][]{ANG99,ILI01} is to adopt 10\%
resonance strength upper limit values for the calculation of the
\textit{recommended} total rates.  ``Lower limits'' or ``upper
limits'' of rates are then derived by completely excluding or by
adopting the full upper limit, respectively, for all resonance
strengths. This procedure is questionable for two reasons. First,
without further knowledge, it is implicitly assumed that the
probability density for the resonance strength is a uniform
distribution extending from zero to the upper limit value. The
implication is that the \textit{mean value} of the resonance strength
amounts to half of its upper limit value. This conclusion contradicts
fundamental nuclear physics, as will be explained below. Second, the
derived ``upper limit'' and ``lower limit'' on the total reaction rate
are usually interpreted as sharp boundaries. This conclusion is also
unphysical, as will be explained below.

The strength of a resonance depends on particle partial widths, which
can be expressed in terms of reduced widths, $\gamma^2$, or,
alternatively, spectroscopic factors, $S$ (see
section~\ref{sec:rates-formalism}). 
These quantities depend on the overlap between the incoming channel
($a+A$) and the compound nucleus final state, which in turn depends on
a nuclear matrix element.
If the nuclear matrix element has contributions from many different
parts of configuration space, and if the signs of these contributions
are random, then the central limit theorem predicts that the
probability density of the transition amplitude will tend toward a
Gaussian distribution centred at zero. The probability density of the
reduced width, representing the \textit{square} of the amplitude, is
then given by a chi-squared distribution with one degree of
freedom. These arguments were first presented by \textcite{POR56} and
this probability density is also known as the Porter-Thomas
distribution. For a particle channel it can be written as
\begin{equation}
  \label{eq:PT}
  f(x) = \frac{c}{\sqrt{\theta^2}}e^{-\theta^2/(2 \hat{\theta}^2)}
\end{equation}
where $c$ is a normalisation constant, $\theta^2$ is the dimensionless
reduced width, and $\hat{\theta}^2$ is the local mean value of the
dimensionless reduced width. The distribution implies that the reduced
width for a given nucleus and set of quantum numbers varies by several
orders of magnitude, with a higher probability the smaller the value
of the reduced width. The Porter-Thomas distribution emerges naturally
from the Gaussian orthogonal ensemble of random matrix theory and is
well established experimentally (see Ref.\ \cite{WEI09} for a recent
review)\footnote{Recently, a high precision study of neutron partial
  widths in plutonium ($A=192, 194$) by \textcite{KOE10} and a
  re-analysis of the Nuclear Data Ensemble ($A=64 - 238$) in Ref.\
  \cite{KOE11} have claimed that the data are not well described by a
  $\chi^2$ distribution with one degree of freedom ($\nu=1$, i.e., a
  Porter-Thomas distribution). They find, depending on the data set
  under consideration, values between $\nu=0.5$ and $\nu=1.2$. These
  new results are controversial and more studies are needed before the
  issue can be settled. It is not clear at present if this controversy
  has any implications for the compound nucleus \nuc{26}{Mg}.}.

The above discussion provides a physically sound method for randomly
sampling reduced widths (or spectroscopic factors) if only an upper
limit value is available. Furthermore, in the present work we
assume a sharp truncation of the Porter-Thomas distribution at the
upper limit value for the dimensionless reduced width,
$\theta^2_{ul}$, that is, we randomly sample over the probability
density
\begin{equation}
  \label{eq:rates-portthom}
  f(\theta) =\left\{\begin{array}{ll}
      \dfrac{c}{\sqrt{\theta^2}}e^{-\theta^2/(2
        \hat{\theta}^2)} &\mbox{ if }\theta^2 \leq \theta_{ul}^2\\
      0 &\mbox{ if }\theta^2>\theta_{ul}^2 \end{array}\right.
\end{equation}
Once dimensionless reduced widths are obtained from sampling according
to equation~(\ref{eq:rates-portthom}), samples of particle partial
widths can be found from equation~(\ref{eq:rates-GammaDefinition}).
Subsequently, samples of resonance strengths can be determined from
equation~(\ref{eq:rates-ResStrength}).

In order to utilise equation~(\ref{eq:rates-portthom}) for Monte Carlo
sampling of $\alpha$-particle partial widths, the mean value of the
dimensionless reduced width, $\hat{\theta}_{\alpha}^2$, must be
known. To this end we considered 360 $\alpha$-particle reduced widths
in the A=20-40 mass region \citep[see][and references
therein]{DRA94}. The distribution is shown in
figure~\ref{fig:SpecFacts} as a black histogram. Binning and fitting
the data to equation~(\ref{eq:PT}) (solid line) results in a best-fit
value of $\hat{\theta}_{\alpha}^2=0.010$, which we adopt in the
present work. It is important to recall the above arguments: the
distribution of reduced widths for a given nucleus, given orbital
angular momentum, given channel spin, etc., is expected to follow a
Porter-Thomas distribution. However, because of the relatively small
sample size of 360 values, we were compelled to fit the entire set by
disregarding differences in nuclear mass number and orbital angular
momentum. For this reason, our derived mean value of 0.010 must be
regarded as preliminary. More reliable estimates of
$\hat{\theta}_{\alpha}^2$ have to await the analysis of a
significantly larger data set of $\alpha$-particle reduced widths when
it becomes available in the future.

From the arguments presented above it should also be clear that the
Porter-Thomas distribution is not expected to represent the reduced
width of all nuclear levels, particularly if the amplitude is
dominated by a few large contributions of configuration space. The
most important example for the latter situation are $\alpha$-cluster
states, which are expected to have relatively large reduced
widths. Indeed, the large reduced width values in
figure~\ref{fig:SpecFacts} that are not described by the Porter-Thomas
distribution (solid line) originate most likely from $\alpha$-cluster
states. Clearly, the nuclear structure of a level in question must be
considered carefully. For this reason, results from $\alpha$-particle
transfer studies are very important. It can be argued that these
measurements populate preferentially $\alpha$-cluster states, with
large reduced widths (or spectroscopic factors), while levels not
populated in $\alpha$-transfer have small reduced widths and,
therefore, are more likely statistical in nature (i.e., described by a
Porter-Thomas distribution). This issue will become important in later
sections.

Once a random sampling of all input parameters has been performed, an
ensemble of reaction rates is obtained. From its probability density
one can extract descriptive statistics (mean, median, variance
etc.). For the recommended reaction rate, we adopt the median value.
The median is a useful statistic because exactly half of the
calculated rates lie above this value and half below. Note that we do
not use the mean value because it is strongly affected by outliers in
the reaction rate distribution.  The low and high reaction rates are
obtained by assuming a 68\% coverage probability. There are several
methods for obtaining these coverage probabilities, such as finding
the coverage that minimises the range of the uncertainties, or one
that is centred on the median.  In the present work, the
$16^{\text{th}}$ to 84$^{\text{th}}$ percentiles of the cumulative
reaction rate distribution are used. We emphasise an important point
regarding reaction rate uncertainties: contrary to previous work, our
``low'' and ``high'' rates do not represent sharp boundaries (i.e., a
probability density of zero outside the boundaries). As with any other
continuous probability density function, these values depend on the
assumed coverage probability, i.e., assuming a larger coverage will
result in a larger uncertainty of the total reaction rate (this is
further illustrated in Figs.~\ref{fig:Uncerts_ag}
and~\ref{fig:Uncerts_an}). The important point here is that the Monte
Carlo sampling results in ``low'' and ``high'' rates for which the
coverage probability can be quantified precisely.

Although a low, high and median rate are useful quantities, they do
not necessarily contain all the information on the rate probability
density.  For application of a reaction rate to nucleosynthesis
calculations, therefore, it is useful to approximate the rate
probability density by a simple analytical approximation. It was shown
in \cite{LON10} that in most (but not all) cases the reaction rate
probability density is well approximated by a lognormal distribution
(equation~\ref{eq:rates-lognormal}). The lognormal parameters $\mu$
and $\sigma$ can be found from the sampled total rates at each
temperature according to
\begin{equation}
  \label{eq:rates-lognormPars}
  \mu = E[\ln(y)],\qquad \sigma^2 = V[\ln(y)]
\end{equation}
where $E[\ln(y)]$ and $V[\ln(y)]$ denote the expectation value and
variance of the natural logarithm of the total rate, $y$,
respectively. A useful measure of the applicability of a lognormal
approximation to the actual sampled distribution is provided by the
Anderson-Darling statistic\footnote{The Anderson-Darling
  statistic~\citep{AND54} is more useful than a $\chi^2$ statistic
  because it does not require binning of the data. The latter usually
  results in a loss of information.}, which is calculated from
\begin{equation}
  \label{eq:rates-AD}
  t_{AD} = -n - \sum_{i=1}^n\frac{2i-1}{n}(\ln F(y_i) +
  \ln\left[1-F(y_{n+1-i})\right]
\end{equation}
where $n$ is the number of samples, $y_i$ are the sampled reaction
rates at a given temperature (arranged in ascending order), and $F$ is
the cumulative distribution of a standard normal function (i.e., a
Gaussian centred at zero). An A-D value greater than unity indicates a
deviation from a lognormal distribution. However, it was found by
\textcite{LON10} that the rate distribution does not \emph{visibly}
deviate from lognormal until A-D exceeds $t_{AD} \approx 30$.  The A-D
statistic is presented in Tabs.~\ref{tab:agRate} and~\ref{tab:anRate}
along with the reaction rates at each temperature in order to provide
a reference to the reader.

\subsection{Extrapolation of Experimental Reaction Rates to Higher Temperatures}
\label{sec-2_4}

Experimental rates usually need to be extrapolated to high
temperatures with the aid of theoretical models because resonances are
only measured up to some finite energy,
$E_{\text{max}}^{\text{exp}}$. If the effective stellar burning energy
window\ \citep{NEW08} extends above this energy, the rate calculated
using the procedure outlined above will become inaccurate. Statistical
nuclear reaction models must, therefore, be used to extrapolate the
experimental rates beyond this temperature. The method used here is
described in detail in Ref.\ \cite{NEW08}. It uses the following strategy:
(i) an {\textit{effective thermonuclear energy range}} (ETER) is
defined using the 8$^{\text{th}}$, 50$^{\text{th}}$, and
92$^{\text{nd}}$ percentiles of the cumulative distribution of
fractional reaction rates (i.e., the relative contribution of single
resonances at temperature $T$ divided by the total reaction rate at
$T$); (ii) the temperature, $T_{\text{match}}$, beyond which the total
rate must be extrapolated is estimated from
\begin{equation}
  \label{eq:HFMatch} E(T_{\text{match}}) + \Delta
  E(T_{\text{match}}) = E^{\text{exp}}_{\text{max}}
\end{equation}
where $\Delta E(T_{\text{match}})$ is the width of the
ETER calculated from the 8$^{\text{th}}$ and 92$^{\text{nd}}$ rate
percentiles.  
We adopt the Hauser-Feshbach rates of Ref.\ \cite{RAU00} for
temperatures beyond $T_{\text{match}}$, normalised to the experimental
rate at $T_{\text{match}}$.

\section{The \Nepa Reactions}
\label{sec:rates}

\subsection{General Aspects}
\label{sec:general-aspects}

The \an ($Q_{\alpha n}=-478.296 (89)$\ keV) and \ag ($Q_{\alpha
  \gamma}=10614.787 (33)$\ keV) reactions are both important in
s-process neutron production. While the \an reaction produces
neutrons, the \ag reaction also influences the neutron flux by
directly competing for available $\alpha$-particles. The rates of both
reactions will therefore be presented here. The centre-of-mass energy
region of interest to the s-process amounts to E$_{cm}=600 \pm 300$\
keV, corresponding to excitation energies of E$_{x}=10900 - 11500$~keV
in the \nuc{26}{Mg} compound nucleus. Note that only states of
``natural'' parity (i.e., $0^+, 1^-, 2^+$, etc.) can be populated via
\Nepa (because both target and projectile have spin-parities of
$0^+$).

Since the early 1980's, several direct measurements were performed of
both reactions close to the energy region of interest\
\citep{WOL89,HAR91,DRO91,DRO93,GIE93,JAE01}. All of these
measurements, with the exception of Ref.\ \cite{GIE93}, were made
using gas targets at the Institut f\"{u}r Strahlenphysik in Stuttgart,
Germany\ \citep[e.g.,][]{HAM98}. The lowest energy resonance measured
in those works is located at $E_{r}^{\text{lab}} \approx 830$\ keV,
near the high energy end of the astrophysically important region. The
structure of the \nuc{26}{Mg} compound nucleus near the
$\alpha$-particle and neutron thresholds has been investigated
previously via neutron capture \citep{SIN74,WEI76}, scattering
\citep{FAG75,MOS76,TAM03}, photoexcitation \citep{BER84,CRA89,SCH09},
transfer \citep{GLA86,YAS90,GIE93,UGA07}, and photoneutron
measurements \citep{BER69}. In particular, the latter study observed
the strong population of a \nuc{26}{Mg} level near \Ex{11150}, with
presumed quantum numbers of J$^{\pi}=1^-$, corresponding to an
expected low-energy resonance at E$_{cm}=450$~keV. It was believed to
have been observed by \textcite{DRO91} and\ \textcite{HAR91} at
\Erlab{630}, but the presumed signal was later shown to be caused by
background from the $^{11}$B($\alpha$,n)$^{14}$N
reaction. Nevertheless, the anticipated contribution from this
low-energy resonance has sensitively influenced all past estimates of
\Nepa reaction rates. For example, it was shown by \textcite{THE00}
that it has a strong impact on s-process nucleosynthesis in massive
stars. However, recent \mggg studies by~\textcite{LON09} demonstrated
unambiguously that this particular level has unnatural parity
(\Jpi{1}{+}) and, therefore, cannot be populated via $\alpha$-particle
capture on \nuc{22}{Ne}.

Studies that provide new experimental information relevant to
$^{22}$Ne$+\alpha$, obtained after the NACRE compilation was
published~\citep{ANG99}, are summarised in
table~\ref{tab:data-since-NACRE}. The goal of the following discussion
is to consider all the available experimental information for states
in \nuc{26}{Mg} of interest to s-process nucleosynthesis and to assign
these levels to corresponding resonances in both \ag and
$^{22}$Ne($\alpha$,n)$^{25}$Mg. This allows for an estimation of the
partial and total resonance widths, resulting in more accurate \Nepa
reaction rates.  A number of levels in \nuc{26}{Mg} near the
$\alpha$-particle and neutron thresholds have unknown spin-parities
and partial widths. These levels have been disregarded in all previous
reaction rate estimates. Since it is not known at present if any of
these are natural parity states and, therefore, may be populated in
$^{22}$Ne$+\alpha$, they cannot be easily included in a Monte Carlo
reaction rate analysis at present. Thus, our strategy is as follows:
we will first derive \Nepa Monte Carlo rates by excluding these levels
of unknown spin-parities. Subsequently, we will investigate their
impact on the total reaction rates {\textit{under the extreme
    assumption that all of these levels possess natural parity}}. As
will be seen below, future measurements of these states are highly
desirable. Throughout the following discussion, energies are presented
in the centre of mass frame unless otherwise stated.

\begin{table*}
  \centering
    \begin{tabular}{cc|l}
      \hline \hline
      Reference    & Reaction Studied              & Comments                                                \\ \hline
      \textcite{JAE01} & \an                           & Resonances between \Erlab{570} and \Erlab{1450} \\
      \textcite{KOE02} & $^{\text{nat}}$Mg(n,$\gamma$)  & $E_{x}$, $J^{\pi}$, $\Gamma_{\gamma}$, $\Gamma_{n}$ for states corresponding to \Erlab{570} to \Erlab{1000} \\
      \textcite{UGA07} & \reaction{22}{Ne}{$^{6}$Li}{d}{26}{Mg} & $J^{\pi}$ and $S_{\alpha}$ for two states below neutron threshold \\
      \textcite{LON09} & \reaction{26}{Mg}{$\Vec{\gamma}$}{$\gamma'$}{26}{Mg} & $E_{x}$ and $J^{\pi}$ for four resonances corresponding to \Erlab{38} to \Erlab{636} \\
      \textcite{DEB10} & \reaction{26}{Mg}{$\Vec{\gamma}$}{$\gamma'$}{26}{Mg} & $\Gamma_{\gamma}$ for four resonances corresponding to \Erlab{38} to \Erlab{636} \\
      \hline \hline
    \end{tabular}
  \caption{\label{tab:data-since-NACRE}New information relevant to the \Nepa reaction rates that has
    become available since the NACRE compilation was published \citep{ANG99}.}
\end{table*}

\subsection{Resonance Strengths}
\label{sec:resonance-strengths}

Directly measured resonance strengths in the \ag reaction in the
energy range of E$_r^{\text{lab}}=830 - 2040$~keV are adopted from
Ref.\ \cite{WOL89}. Direct measurements of resonances in the \an
reaction at energies of E$_r^{\text{lab}}=830 - 2040$~keV are reported
in Refs.\ \cite{WOL89,HAR91,DRO93,GIE93,JAE01}. Note, however, that
the \an resonance strengths from the different measurements disagree
by up to a factor of 5 (i.e., a deviation well outside the quoted
uncertainties). Clearly, adopting a simple weighted average value
would not account for the unknown systematic bias present in the
data. To alleviate this problem, we adopt the method of Ref.\
\cite{WIE11}, previously applied to account for unknown systematic
uncertainties in neutron-lifetime measurements. This method follows a
similar procedure for characterising unknown systematic uncertainties
as that presented in Ref.\ \cite{DES04}. It assumes that all the
reported strength values of a given resonance have the same, unknown,
systematic error, $\sigma_u$, which can be summed in quadrature with
the reported uncertainties. Hence for each reported uncertainty of
data set $i$, $\sigma_i$, an inflated uncertainty, $\sigma_i'$, is
obtained via
\begin{equation}
  \label{eq:inflated-uncertainty}
  \sigma_i' = \sqrt{\sigma_u^2 + \sigma_i^2}
\end{equation}
From the inflated uncertainties, the weighted average of the resonance
strengths, $\omega \gamma_i$, is obtained in the usual manner,
\begin{equation}
  \label{eq:weightedaverage}
  \overline{\omega \gamma} = \frac{\sum_i \omega
    \gamma_i/\sigma^{'2}_i}{\sum_i 1/\sigma^{'2}_i} \qquad
  \overline{\sigma} = \sqrt{\frac{1}{\sum_i 1/\sigma^{'2}_i}}
\end{equation}
The unknown value of $\sigma_u$ is adjusted numerically until the
reduced chi-squared, $\chi^2/\nu$, becomes equal to unity.
\begin{equation}
  \label{eq:chi2}
  \frac{\chi^2}{\nu} = \frac{1}{n-1}\sum_i{\frac{\left(\omega \gamma_i - \overline{\omega
          \gamma}\right)^2}{\sigma^{'2}_i}}
\end{equation}
where $\nu$ is the degree of freedom (i.e., $\nu = n-1$, with $n$
equal to the number of measurements).

Application of this method has two consequences compared to
calculating the weighted average of the reported resonance strength
values: (i) the uncertainty of the resonance strength,
$\overline{\sigma}$, will be larger, reflecting the fact that the
systematic shift in the data is of unknown nature; and (ii) strength
values with small reported uncertainties will carry less
weight. Consider as an example the lowest-lying observed resonance in
the \an reaction, located at \Erlab{831}. The measurements reported in
Refs.\ \cite{HAR91,DRO93,GIE93,JAE01} yield for the resonance strength
a (standard) weighted average of $\omega \gamma = 1.2 (1) \times
10^{-4}$~eV, with $\chi^2/\nu = 2.9$, indicating poor agreement
between the individual measurements. On the other hand, the inflated
weighted average value is $\omega \gamma = 1.4 (3) \times
10^{-4}$~eV. We applied the inflated weighted average method to all
resonances in the energy region E$_r^{\text{lab}}=830 -
1495$~keV. Above this energy range, we used the (standard) weighted
average because the different data sets are in considerably better
agreement.

From the measured \an and \ag strengths of a given resonance, the
neutron and $\alpha$-particle partial widths can be found if the
$\gamma$-ray partial width can be estimated. This information allows
for integrating the resonance cross section numerically, according to
equation~(\ref{eq:rates-rr-resonance}), which is more reliable than
adopting the narrow resonance approximation,
equation~(\ref{eq:rates-narrowrate}). 
Because of Coulomb barrier penetrability arguments, the neutron width
is expected to dominate the total width of the resonances important
for s-process nucleosynthesis (i.e., $\Gamma_n \approx \Gamma$). Thus,
in most (but not all) cases, the neutron width exceeds the
$\alpha$-particle width for a given state substantially and we can use
the following approximations to determine the $\alpha$-particle width
from measured resonance strengths. For the \ag reaction, the
$\alpha$-particle partial width can be found by assuming a reasonable
average value for the \gray\ partial width of $\Gamma_{\gamma} \approx
3$\ eV\ \citep{KOE02}. We investigated the effect of this
  choice on the reaction rates and the exact average value of
  $\Gamma_{\gamma}$ was found to be relatively unimportant. The
$\alpha$-particle partial width can then be found (for $\Gamma_n
\approx \Gamma$) from
\begin{equation}
  \omega \gamma_{\alpha\gamma} = \omega \frac{\Gamma_{\alpha}
    \Gamma_{\gamma}}{\Gamma_{\alpha}+\Gamma_{\gamma}+\Gamma_{n}}, \qquad
  \Gamma_{\alpha}= \frac{\omega
    \gamma_{\alpha\gamma}}{\omega}\frac{\Gamma}{3\ eV} 
\end{equation}
For the \an reaction, the $\alpha$-particle partial width can be
calculated from:
\begin{equation} 
  \omega \gamma_{{\alpha}n} =
  \omega \frac{\Gamma_{\alpha}
    \Gamma_{n}}{\Gamma_{\alpha}+\Gamma_{\gamma}+\Gamma_{n}}, \qquad
  \Gamma_{\alpha}= \frac{\omega \gamma_{{\alpha}n}}{\omega} \label{eqn:approxGa}
\end{equation}

\subsection{Spectroscopic Factors}
\label{sec:spectro-factors}

Alpha-particle spectroscopic factors for levels near the
$\alpha$-particle and neutron thresholds in \nuc{26}{Mg} have been
obtained from \reaction{22}{Ne}{$^6$Li}{d}{26}{Mg} transfer studies by
Refs.\ \cite{GIE93} and \cite{UGA07}.  The spectroscopic factors
derived from the ($^6$Li,d) transfer data are important because they
allow for an estimate of the $\alpha$-particle partial width,
$\Gamma_{\alpha}$, of \Nepa resonances via
equations~(\ref{eq:S-reduced-width-relation1}),
(\ref{eq:S-reduced-width-relation2}),
and~(\ref{eq:rates-GammaDefinition}).

\begin{figure*} 
  \begin{center}
    \includegraphics[width=0.6\textwidth]{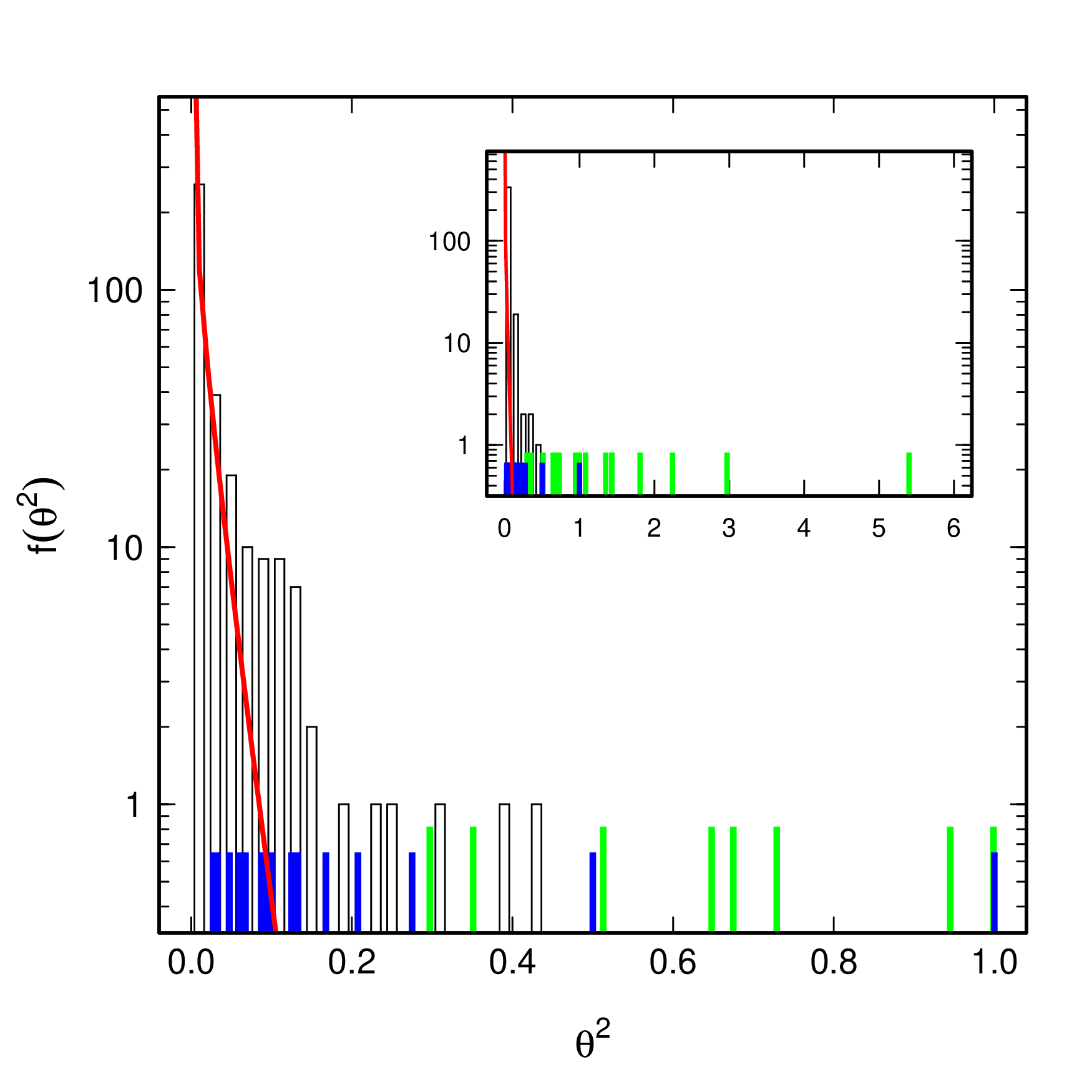} 
    \caption[Spectroscopic Factors]{\label{fig:SpecFacts}(Colour
      online) Dimensionless reduced $\alpha$-particle widths of
      unbound states from Ref.\ \cite{DRA94}, and references therein
      \citep[see also,][]{LON10}. Also plotted is the Porter-Thomas
      distribution that best fits these data at small values. It is
      apparent from the figure that states with large
      $\alpha$-particle spectroscopic factors are not represented by
      the Porter-Thomas distribution. These levels most likely have an
      $\alpha$-particle cluster structure and would be populated
      preferentially in transfer measurements, such as the
      ($^{6}$Li,d) measurements of Refs.\ \cite{GIE93} and
      \cite{UGA07}. Also shown in green and blue are the normalised
      spectroscopic factors measured by Ref.\ \cite{GIE93}. The values
      normalised using the \Erlab{1434} resonance are shown in blue,
      while those normalised to the \Erlab{831} resonance are shown in
      green. Clearly, the normalisations are vastly different, and the
      spectroscopic factors obtained using the \Erlab{831} resonance
      as a normalisation reference appear to be too high, as shown in
      the figure inset, which displays the same information but on an
      expanded scale. See text for more detail.}
  \end{center}
\end{figure*}

Numerous studies have shown that $\alpha$-transfer measurements are
very useful for measuring {\textit{relative}} spectroscopic factors,
but are not sufficiently accurate for predicting {\textit{absolute}}
values. For this reason, the measured spectroscopic factors are
frequently scaled relative to resonances with well-known partial
widths (note that this is an approximation equivalent to assuming that
$\theta^2=S$ in Sec.\ \ref{sec:formalism}). For example,
\textcite{GIE93} scaled their spectroscopic factors relative to the
\Erlab{831} (\Jpi{2}{+}) resonance in
\reaction{22}{Ne}{$\alpha$}{n}{25}{Mg}. Our best value for the
($\alpha$,n) resonance strength is $\omega \gamma_{\alpha n} = 1.4 (3)
\times 10^{-4}$~eV (see Tab.~\ref{tab:an_direct}). Since for this
low-energy resonance it can be safely assumed that $\Gamma \approx
\Gamma_n$, a spectroscopic factor of $S_{\alpha}^{(\alpha,n)}=0.98$ is
obtained from equations~(\ref{eq:rates-ResStrength})
and~(\ref{eq:rates-GammaDefinition}). Surprisingly, this value is a
factor of 27 larger than the transfer value extracted by Ref.\
\cite{GIE93}, $S_{\alpha}^{(^6\text{Li},d)} = 0.037$. Renormalisation
of all measured ($^6$Li,d) spectroscopic factors to the ($\alpha$,n)
spectroscopic factor of the \Erlab{831} resonance results in the
values shown in green in Fig.~\ref{fig:SpecFacts}. It is certainly
remarkable that all levels observed by Ref.\ \cite{GIE93} should have
dimensionless reduced $\alpha$-particle widths far larger in value
than the Porter-Thomas prediction. Additionally, several of these
levels exhibit dimensionless reduced widths near or exceeding the
Wigner limit, even if one accounts for the difference between
$S_{\alpha}$ and $\theta^2_{\alpha}$ (Sec.~\ref{sec:rates-formalism}).

However, there is no compelling reason why the \Erlab{831} resonance
should be singled out for the normalisation procedure, other than it
being the lowest-lying observed resonance. 
For example, one may consider another well-known resonance, located at
\Erlab{1434} (\Jpi{2}{+}). From its measured \an resonance strength,
an $\alpha$-particle spectroscopic factor of $S_{\alpha}^{(\alpha,n)}
= 0.27$ is obtained. The $\alpha$-particle transfer value for the
corresponding levels, measured by Ref.\ \cite{GIE93}, amounts to
$S_{\alpha}^{(^6\text{Li},d)} = 0.11$.  These two values differ by a
factor of 2.5, and thus are much closer in agreement than the results
for the \Erlab{831} resonance. Normalisation of all measured relative
($^6$Li,d) spectroscopic factors to the ($\alpha$,n) spectroscopic
factor of the \Erlab{1434} resonance results in the values shown in
blue in Fig.~\ref{fig:SpecFacts}. It is evident that these normalised
values are in far better agreement with the Porter-Thomas distribution
than the results obtained when scaling spectroscopic factors relative
to the \Erlab{831} resonance. In addition, by using the \Erlab{1434}
resonance normalisation, all of the resulting dimensionless reduced
widths now have values less than the Wigner limit, making them more
believable. Since we feel it is more reasonable to scale the
($^6$Li,d) spectroscopic factors using the \Erlab{1434} resonance
instead of the \Erlab{831} resonance, we adopt the reduced widths
shown in blue in Fig.~\ref{fig:SpecFacts} for calculating the \Nepa
rates. Note that the {\textit{relative}} spectroscopic factors
obtained by Ref.\ \cite{GIE93} have been used at face value by Refs.\
\cite{KOE02,KAR06} in their reaction rate calculations.  Clearly, this
issue needs to be resolved in future work.

\section{Information on Specific \nuc{26}{Mg} Levels}
\label{sec:levels}

\Exb{10693} (\Erlabb{92}; \mbox{\Jpib{4}{+}}). An excited state near
this energy has been observed by \textcite{GLA86} at E$_{x}$=10695(2)\
keV, \textcite{GIE93} at E$_{x}$=10694(20)\ keV, and \textcite{MOS76}
at E$_{x}$=10689(3)\ keV. A weighted average of these excitation
energies is used in the present work. The \Jpi{4}{+} assignment was
made by considering the decay scheme of this state as observed by
Ref.\ \cite{GLA86} and that the state most likely has natural
parity. The $\alpha$-particle spectroscopic factor for this state from
Ref.\ \cite{GIE93}, after normalisation, is $S_{\alpha}=0.059$.

\Exb{10806} (\Erlabb{226}; \mbox{\Jpib{1}{-}}). This state was seen
previously in $^{22}$Ne($^{6}$Li,d)$^{26}$Mg measurements by
\textcite{UGA07} at E$_{x}$=10808 (20)\ keV, and in
$^{25}$Mg(n$_{t}$,$\gamma$)$^{26}$Mg measurements (thermal neutron
capture) by \textcite{WAL92} at E$_{x}$=10805.9 (4)\ keV. A recent
experiment assigned a spin-parity of \Jpi{1}{-}\ \citep{LON09}. The
adopted excitation energy is the weighted average of these
results. The $\alpha$-particle spectroscopic factor from Ref.\
\cite{UGA07}, after normalisation, amounts to $S_{\alpha}=0.048$.

\Exb{10943} (\Erlabb{388}; $\mathbf{J^{\pi}=(5^{-}-7^{-})}$). An
excited state at this energy has been observed by
\textcite{GLA86}. The observed decay scheme restricts the quantum
numbers, using the dipole-or-E2 rule of Ref.\ \cite{END90}, to
$J^{\pi}=(5^{\pm}-7^{-})$. A $^{22}$Ne($^{6}$Li,d)$^{26}$Mg transfer
measurement populated a state at E$_{x}$=10953 (25)\ keV, but did not
obtain the quantum numbers other than to report that it most likely
had natural parity\ \citep{UGA07}. The combined quantum number
assignment is therefore $J^{\pi}=(5^{-}-7^{-})$. This state is treated
here as part of a doublet with the \Ex{10949} state. Note that in the
reaction rate calculations of \textcite{KAR06}, this state was
incorrectly assigned spin-parity values of $J^{\pi}=2^+, 3^-$.

\Exb{10949} (\Erlabb{395}; \mbox{\Jpib{1}{-}}). This state has been
observed previously in $^{26}$Mg(p,p$'$)$^{26}$Mg measurements by
\textcite{MOS76} at E$_{x} = 10950 (3)$\ keV. The (p,p$'$)
measurements suggest a $J^{\pi}=1^{-}$ assignment, which agrees with
the \mggg result of \textcite{LON09}. It is unclear whether
\textcite{UGA07} observed this state or the one at E$_{x}$=10943\
keV. Therefore, in the present analysis, the normalised spectroscopic
factor of $S=7 \times 10^{-3}$ reported in Ref.\ \cite{UGA07} is
treated as an upper limit for both states at \Ex{10943} and
\Ex{10949}.

\Exb{11112} (\Erlabb{587}; \mbox{\Jpib{2}{+}}). The state is located
above the neutron threshold. It has been observed previously in a
\reaction{25}{Mg}{n}{$\gamma$}{26}{Mg} experiment\ \citep{WEI76,KOE02}
and was assigned a spin-parity of \Jpi{2}{+}. 

\Exb{11154} (\Erlabb{637}; \mbox{\Jpib{1}{+}}). 
A state at this energy has been observed by \textcite{FAG75},
\textcite{WEI76}, \textcite{CRA89}, \textcite{YAS90},
\textcite{KOE02}, \textcite{TAM03}, and
\textcite{SCH09}. Additionally, an excited state at this energy was
strongly populated by the photoneutron experiment of \textcite{BER69},
who predicted a \Jpi{1}{-} assignment. As a result of this prediction,
several studies have searched for a resonance corresponding to this
energy~\citep{HAR91,DRO91,DRO93,GIE93,JAE01,UGA07}. Of these studies,
a presumed resonance was reported by Refs.\ \cite{HAR91,DRO91}, but
later proven to be caused by beam induced background
\citep{DRO93}. Recently, however, a \mggg experiment \citep{LON09}
showed unambiguously that the spin-parity of this state amounts to
\Jpi{1}{+} and, therefore, cannot contribute to the \Nepa reactions
rates. A more detailed discussion of this state is presented in
section~\ref{sec:general-aspects}.

\textbf{E$\mathbf{_{x}}$=11163-11326\ keV
  (E$\mathbf{_{r}^{\textbf{lab}}=648-840}$~keV)}. Excitation energies
were taken as weighted averages of \textcite{MOS76}, \textcite{GLA86},
and \textcite{KOE02}. Quantum numbers, neutron and $\gamma$-ray
partial widths were all adopted from Ref.\ \cite{KOE02}. Since no
$\alpha$-particle partial widths have been measured for these states,
upper limits were derived either from the data presented by Ref.\
\cite{WOL89}, or adopted from the maximum theoretically allowed
values, depending on which was smaller.

\Exb{11318} (\Erlabb{831}; \mbox{\Jpib{2}{+}}). \textcite{KOE02}
argued that this state cannot correspond to both the resonance
observed by \textcite{JAE01} at E$_{r}$=832 (2)\ keV in \an and by
\textcite{WOL89} at E$_{r}^{lab}=828 (5)$\ keV in
\reaction{22}{Ne}{$\alpha$}{$\gamma$}{26}{Mg}, because the implied
value of $\Gamma_{\gamma}=76$~eV would be far larger than the average
$\gamma$-ray partial width ($\Gamma_{\gamma}=3$~eV) in this energy
range. However, this conclusion is questionable considering the large
uncertainty, $\Gamma_{\gamma}=76 (53)$~eV, when the $\gamma$-ray
partial width is derived from the measured values of $\omega
\gamma_{\alpha \gamma}$, $\omega \gamma_{\alpha n}$, and
$\Gamma$. Clearly, the deviation from the average in this energy range
amounts to only $1.4\sigma$.

Since it cannot be decided at present if the ($\alpha$,n) and
($\alpha,\gamma$) resonances correspond to the same \nuc{26}{Mg} level
or not, the partial widths cannot be derived unambiguously from the
measured resonance strengths and total width. Thus we assumed that the
($\alpha$,n) and ($\alpha,\gamma$) resonances are ``narrow'', i.e., we
employed equation~(\ref{eq:rates-narrowrate}) instead of
equation~(\ref{eq:rates-rr-resonance}) in our rate calculations. The
strength reported by Ref.\ \cite{WOL89} is used for the \ag resonance, while
the inflated weighted average (see
section~\ref{sec:resonance-strengths}) is adopted for the \an
resonance, resulting in a strength of $\omega \gamma_{(\alpha,n)} =
1.4 (3) \times 10^{-4}$~eV.

\textbf{E$\mathbf{_{x}}$=11328--11425\ keV
  (E$\mathbf{_{r}^{\textbf{lab}}=843-957}$~keV)}. Excitation energies,
quantum numbers, neutron, and \gray\ widths for these levels are
adopted from Refs.\ \cite{WEI76}, and \cite{KOE02}. No
$\alpha$-particle widths have been measured for these states, and thus
upper limits have been adopted from either the data presented by
Refs.\ \cite{WOL89} and \cite{JAE01}, or from the maximum
theoretically allowed values, depending on which was smaller.

\textbf{E$\mathbf{_{x}>11441}$\ keV
  (E$\mathbf{_{r}^{\textbf{lab}}>976}$~keV)}. Resonances corresponding
to excited states above \Ex{11441} have been measured directly\
\citep{WOL89,HAR91,DRO91,DRO93,GIE93,JAE01}. In order to take the
widths of wide resonances into account, the neutron and \gray\ partial
widths (and quantum numbers) measured by Refs.\ \cite{WEI76} and
\cite{KOE02} have been used when available. The inflated weighted
average method (see section~\ref{sec:resonance-strengths}) is used to
combine the different \an strengths for resonances below \Erlab{1434},
while standard weighted averages are used above this energy. Since the
\ag resonances measured in Ref.\ \cite{WOL89} cannot be assigned
unambiguously to corresponding \an resonances, all of these resonances
were treated as independent and narrow. The quantum numbers of \an
resonances located above \Erlab{1530} are adopted from Ref.\
\cite{WOL89} when not available otherwise.

\section{Reaction Rates for \Nepa}
\label{sec:rates-results}

The resonance properties used to calculate the rates for both the \ag
and the \an reactions are presented in
Tabs. \ref{tab:ag_direct}~--~\ref{tab:an_upperlims}. For more detailed
information on level properties, see Ref.\ \cite{LON10T}. Separate
tables are used to list resonances with measured partial widths and
those which possess only an upper limit for the $\alpha$-particle
width but have known neutron and $\gamma$-ray widths.

\begin{table*} 
  \centering
    \begin{tabular}{cccr@{$\times$}l|r@{$\times$}lcr@{$\times$}lr@{$\times$}l|c}
      \hline \hline &                                                   &                  & \multicolumn{2}{c|}{ }                   & \multicolumn{7}{c|}{Partial Widths (eV)} &                                                                                                                                                                                                \\ 
      E$_{x}$ (keV) & \multicolumn{1}{c}{E$_{r}^{\textrm{lab}}$  (keV)} & J$^{\pi}$ $^{c}$ & \multicolumn{2}{c|}{$\omega\gamma$ (eV)} & \multicolumn{2}{c}{$\Gamma_{\alpha}$}    & \multicolumn{1}{c}{$\Gamma_{\gamma}$$^{a}$} & \multicolumn{2}{c}{$\Gamma_{n}$} & \multicolumn{2}{c|}{$\Gamma$} & Int                                                                           \\ \hline 
      10693         & 93 (2)                                            & $4^+$            & \multicolumn{2}{c|}{- - - -}             & $3.5 (18)$                               & $10^{-46}$                                  & 3.0 (15)                         & \multicolumn{2}{c}{- - - -}   & \multicolumn{2}{c|}{3.0 (15)} &                                               \\
      11315         & 828 (5)                                           & $2^{+}$          & $3.6 (4)$                                & $10^{-5}$                                & \multicolumn{2}{c}{ - - - - }               & - - - -                          & \multicolumn{2}{c}{ - - - - } & \multicolumn{2}{c|}{ - - - - }  &                                   \\     
      11441         & 976.39 (23)                                       & $4^{+}$          & \multicolumn{2}{c|}{- - - -}             & $4.3 (11)$                               & $10^{-6}$ $^{b}$                            & 3.0 (15)                         & 1.47 (8)                      & $10^3$                        & 1.47 (8)  & $10^3$    & \checkmark            \\ 
      11465         & 1005.23 (25)                                      & $5^{-}$          & \multicolumn{2}{c|}{- - - -}             & $5.0 (15)$                               & $10^{-6}$ $^{b}$                            & 3.0 (15)                         & 6.55 (9)                      & $10^3$                        & 6.55 (9)  & $10^3$    & \checkmark            \\ 
      11508         & 1055.9  (11)                                      & $1^{-}$          & \multicolumn{2}{c|}{- - - -}             & $1.2 (2)$                                & $10^{-4}$ $^{b}$                            & 3.0 (15)                         & 1.27 (25)                     & $10^{4}$                      & 1.27 (25) & $10^{4}$  & \checkmark            \\ 
      11526         & 1075.5 (18)                                       & $1^{-}$          & \multicolumn{2}{c|}{- - - -}             & $4.3 (11)$                               & $10^{-4}$ $^{b}$                            & 3.0 (15)                         & 1.8   (9)                     & $10^{3}$                      & 1.8 (9)   & $10^{3}$  & \checkmark            \\ 
      11630         & 1202.3 (17)                                       & $1^{-}$          & \multicolumn{2}{c|}{- - - -}             & $2.4 (5)$                                & $10^{-3}$ $^{b}$                            & 3.0 (15)                         & 1.35 (17)                     & $10^{4}$                      & 1.35 (17) & $10^{4}$  & \checkmark            \\ 
      11748         & 1345 (7)                                          & $1^{-}$          & \multicolumn{2}{c|}{- - - -}             & $2.0 (3)$                                & $10^{-2}$ $^{b}$                            & 3.0 (15)                         & 6.4 (9)                       & $10^{4}$                      & 6.4 (9)   & $10^{4}$  & \checkmark            \\ 
      11787         & 1386 (3)                                          & $1^{-}$          & \multicolumn{2}{c|}{- - - -}             & $8 (3)$                                  & $10^{-3}$ $^{b}$                            & 3.0 (15)                         & 2.45 (24)                     & $10^{4}$                      & 2.45 (24) & $10^{4}$  & \checkmark            \\ 
      11828         & 1433.7 (12)                                       & $2^{+}$          & $2.5 (3)$                                & $10^{-4}$                                & $1.8 (10) $                                 & $10^{-1}$                        & 3.0 (15)                      & 1.10 (25)                     & $10^3$    & 1.10 (25) & $10^3$   & \checkmark \\ 
      11895         & 1513 (5)                                          & $1^{-}$          & $2.0 (2)$                                & $10^{-3}$                                & \multicolumn{2}{c}{ - - - - }               & - - - -                          & \multicolumn{2}{c}{ - - - - } & \multicolumn{2}{c|}{$< 3000$} &                                               \\
      11912         & 1533 (3)                                          & $1^{-},2^+$      & $3.4 (4)$                                & $10^{-3}$                                & $1.9 (8) $                                  & $10^{+0}$                        & 3.0 (15)                      & 5 (2)                         & $10^{3}$  & 5 (2)     & $10^{3}$ & \checkmark \\ 
      11953         & 1582 (3)                                          & $2^+,3^{-},4^+$  & $3.4 (4)$                                & $10^{-3}$                                & $3.2 (17) $                                    & $10^{-1}$                        & 3.0 (15)                      & 2 (1)                         & $10^{3}$  & 2 (1)     & $10^{3}$ & \checkmark \\ 
      12051         & 1698 (3)                                          & $2^+,3^{-}$      & $6.0 (7)$                                & $10^{-3}$                                & $1.1 (3) $                                    & $10^{-1}$                        & 3.0 (15)                      & 4 (1)                         & $10^{3}$  & 4 (1)     & $10^{3}$ & \checkmark \\ 
      12140         & 1802 (3)                                          & $1^{-}$          & $1.0 (2)$                                & $10^{-3}$                                & $1.7 (5) $                                  & $10^{+0}$                        & 3.0 (15)                      & 15 (2)                        & $10^{3}$  & 15 (2)    & $10^{3}$ & \checkmark \\ 
      12184         & 1855 (8)                                          & ($0^{+}$)        & $1.1 (2)$                                & $10^{-3}$                                & $1.21 (29) $                                & $10^{+1}$                        & 3.0 (15)                      & 33 (5)                        & $10^{3}$  & 33 (5)    & $10^{3}$ & \checkmark \\     
      12273         & 1960 (8)                                          & ($0^{+}$)        & $8.9 (1)$                                & $10^{-3}$                                & $2.2 (4) $                                  & $10^{+2}$                        & 3.0 (15)                      & 73 (9)                       & $10^{3}$  & 73 (9)   & $10^{3}$ & \checkmark \\ 
      12343         & 2043 (5)                                          & $0^{+}$          & $5.4 (7)$                                & $10^{-2}$                                & $6.3 (12) $                                 & $10^{+2}$                        & 3.0 (15)                      & 35 (5)                        & $10^{3}$  & 35 (5)    & $10^{3}$ & \checkmark \\ 
      \hline \hline 
    \end{tabular} 
    \footnotetext{Average value from Ref.\ \cite{KOE02}}
    \footnotetext{From $^{22}$Ne($\alpha$,n)$^{25}$Mg
      measurements (see equation~(\ref{eqn:approxGa}))}
    \footnotetext{Detailed discussion on quantum number
      assignments can be found in
      section~\ref{sec:resonance-strengths}.}
  \caption{\label{tab:ag_direct}Resonances of \ag with known $\alpha$-particle
    partial widths or resonance strengths. Total widths are from
    Ref.\ \cite{WOL89} for resonances above \Erlab{1533}. For lower-lying
    resonances, total widths are adopted from Ref.\ \cite{JAE01} and
    Ref.\ \cite{KOE02}. Ambiguous spin-parities (i.e., those not based on
    strong arguments) are placed in parentheses, according to the
    guidelines in Ref\ \cite{END90}. The last column, labelled ``Int''
    indicates those resonances for which sufficient information is
    available in order to integrate their reaction rate contribution
    numerically, according to
    equation~(\ref{eq:rates-rr-resonance}).}
\end{table*}

\begin{table*}[!ht]
  \begin{center}

      \begin{tabular}{cccr@{$\times$}l|cr@{$\times$}lccc|c}
        \hline \hline 
        &                                                &               & \multicolumn{2}{c|}{ }                                  & \multicolumn{6}{c|}{Partial Widths (eV)}          &                                                                                                                                                                      \\ 
        E$_{x}$ (keV) & \multicolumn{1}{c}{$E_{r}^{\text{lab}}$ (keV)} & J$^{\pi}$     & \multicolumn{2}{c|}{$\omega\gamma_{\mathrm{UL}}$ (eV )} & S$_{\alpha,\mathrm{UL}}$ &\multicolumn{2}{c}{$\Gamma_{\alpha,\mathrm{UL}}$} & \multicolumn{1}{c}{$\Gamma_{\gamma}$} & \multicolumn{1}{c}{$\Gamma_{n}$} & \multicolumn{1}{c|}{$\Gamma$} & Int                                                       \\ \hline 
        10806         & 225.9 (5)               & $1^-$         & \multicolumn{2}{c|}{- - - -}                                                                             & 4.8$\times 10^{-2}$                        & $3.2$                                 & $10^{-23}$                       & $0.72 (18)$                   & - - - -      & $0.72 (18)$   &                            \\
        10943         & 388 (2)                 & $(5^- - 7^-)$ & \multicolumn{2}{c|}{- - - -}                                                                             & 7$\times 10^{-3}$                          & $1.5$                                 & $10^{-19}$                       & $3.0 (15)$                    & - - - -      & $3.0 (15)$    &                            \\
        10949         & 395.15 (18)             & $1^-$         & \multicolumn{2}{c|}{- - - -}                                                                             & 7$\times 10^{-3}$                          & $2.9$                                 & $10^{-15}$                        & $1.9 (3)$                     & - - - -      & $1.9 (3)$     &                            \\
        11112         & 587.90 (10)             & $2^{+}$       & $3.7$                                                                                  & $10^{-08}$      & 1.00                                       & $7.7$                            & $10^{-09}$                    & $1.73 (3)$   & $2578 (240))$ & $2580 (240) $ & \checkmark \\     
        11163         & 647.93 (11)             & $2^{+}$       & $4.3$                                                                                  & $10^{-07}$      & 1.00                                         & $8.7$                            & $10^{-08}$                    & $4.56 (29) $ & $4640 (100) $ & $4650 (100)$  & \checkmark \\ 
        11171         & 657.53 (19)             & $(2^{+})$     & $6.2$                                                                                  & $10^{-07}$      & 1.00                                         & $1.3$                            & $10^{-07}$                    & $3.0 (15)$   & $1.44 (16) $  & $4.4 (15) $   &            \\ 
        11183         & 671.70 (21)             & $(1^{-})$     & $1.0$                                                                                  & $10^{-06}$      & 1.00                                         & $2.1$                            & $10^{-07}$                    & $3.0 (15)$   & $0.54 (9) $   & $3.5 (15) $   &            \\ 
        11243         & 742.81 (12)             & $2^{(-)}$     & $4.7$                                                                                  & $10^{-06}$      & 0.44                                         & $9.5$                            & $10^{-07}$                    & $7.4 (6) $   & $4510 (110) $ & $4520 (110)$  & \checkmark \\ 
        11274         & 779.32 (14)             & $(2)^{+}$     & $4.9$                                                                                  & $10^{-06}$      & 0.15                                         & $1.0$                            & $10^{-06}$                    & $3.2 (4) $   & $540 (50) $   & $540 (50) $   & \checkmark \\ 
        11280         & 786.17 (13)             & $4^{(-)}$     & $8.2$                                                                                  & $10^{-07}$      & 1.00                                         & $9.2$                            & $10^{-08}$                    & $0.59 (24) $ & $1510 (30) $  & $1510 (30) $  & \checkmark \\ 
        11286         & 792.90 (15)             & $1^{-}$       & $5.0$                                                                                  & $10^{-06}$      & 0.05                                         & $1.7$                            & $10^{-06}$                    & $0.8 (5) $   & $1260 (100) $ & $1260 (100)$  & \checkmark \\ 
        11286         & 793.83 (14)             & $(2^{+})$     & $5.0$                                                                                  & $10^{-06}$      & 0.11                                         & $1.0$                            & $10^{-06}$                    & $4.3 (6) $   & $12.8 (6) $   & $17.1 (60) $  & \checkmark \\ 
        11289         & 797.10 (29)             & $(2^{-})$     & $5.1$                                                                                  & $10^{-06}$      & 0.10                                         & $1.0$                            & $10^{-06}$                    & $3.0 (15)$   & $1.5 (5) $    & $4.5 (16) $   &            \\ 
        11296         & 805.19 (16)             & $(3^{-})$     & $5.1$                                                                                  & $10^{-06}$      & 0.39                                         & $7.4$                            & $10^{-07}$                    & $3.3 (7) $   & $8060 (120) $ & $8060 (120)$  & \checkmark \\ 
        11311         & 822.6 (4)               & $(1^{-})$     & $5.2$                                                                                  & $10^{-06}$      & 0.02                                         & $1.8$                            & $10^{-06}$                    & $3.0 (15)$   & $1.1 (4) $    & $4.1 (16) $   &            \\ 
        11326         & 840.8 (6)               & $(1^{-})$     & $5.4$                                                                                  & $10^{-06}$      & 0.01                                         & $1.8$                            & $10^{-06}$                    & $3.0 (15)$   & $0.6 (3) $    & $3.6 (15) $   &            \\ 
        11328         & 843.24 (17)             & $1^{-}$       & $5.4$                                                                                  & $10^{-06}$      & 0.01                                         & $1.8$                            & $10^{-06}$                    & $3.6 (5) $   & $420 (90) $   & $420 (90) $   & \checkmark \\ 
        11329         & 844.4 (6)               & $(1^{-})$     & $5.4$                                                                                  & $10^{-06}$      & 0.01                                         & $1.8$                            & $10^{-06}$                    & $3.0 (15)$   & $2.8 (10) $   & $5.8 (18) $   &            \\ 
        11337         & 853.6 (7)               & $(1^{-})$     & $5.4$                                                                                  & $10^{-06}$      & 0.01                                         & $1.8$                            & $10^{-06}$                    & $3.0 (15)$   & $1.4 (6) $    & $4.4 (18) $   &            \\ 
        11344         & 861.86 (18)             & $(2^{+})$     & $5.5$                                                                                  & $10^{-06}$      & 0.02                                         & $1.1$                            & $10^{-06}$                    & $1.18 (27) $ & $150 (40) $   & $150 (40) $   & \checkmark \\ 
        11345         & 862.91 (19)             & $4^{(-)}$     & $5.5$                                                                                  & $10^{-06}$      & 0.87                                         & $6.2$                            & $10^{-07}$                    & $1.8 (4) $   & $4130 (190) $ & $4130 (190)$  & \checkmark \\ 
        11393         & 919.34 (19)             & $5^{(+)}$     & $1.6$                                                                                  & $10^{-06}$      & 1.00                                         & $1.5$                            & $10^{-07}$                    & $3.0 (15)$   & $290 (19) $   & $290 (19) $   & \checkmark \\  
        \hline \hline 
      \end{tabular} 
    \caption{\label{tab:ag_upperlims}Properties of Unobserved
      Resonances in \ag. For these resonances, only upper limits of
      the resonance strength and/or the $\alpha$-particle
      spectroscopic factor are available at present. The $\gamma$-ray
      and neutron partial widths are taken from the R-matrix fit of
      Ref.\ \cite{KOE02}.  Quantum numbers for states below \Ex{11163}
      are discussed in Sec.\ \ref{sec:levels}. All other quantum
      numbers are adopted from Ref.\ \cite{KOE02}. When a range of
      quantum numbers is allowed, the upper limit of the
      $\alpha$-particle width is calculated assuming the lowest
      possible orbital angular momentum transfer. The upper limit
      $\alpha$-particle spectroscopic factors adopted ($S_{\alpha,
        \mathrm{UL}}$) are also listed for completeness.}
  \end{center} 
\end{table*}

\begin{table*}[!ht]
  \begin{center}
    { \addtolength{\tabcolsep}{2pt}
      \addtolength{\extrarowheight}{2pt}

        \begin{tabular}{cccr@{$\times$}l|r@{$\times$}lcr@{$\times$}lr@{$\times$}l|c}
          \hline \hline 
          &                                                  &                     & \multicolumn{2}{c|}{ }                   & \multicolumn{7}{c|}{Partial Widths (eV)}    &                                                                                                                                                                                                    \\ 
          E$_{x}$ (keV) & \multicolumn{1}{c}{E$_{r}^{\textrm{lab}}$ (keV)} & J$^{\pi}$ $^{a}$    & \multicolumn{2}{c|}{$\omega\gamma$ (eV)} & \multicolumn{2}{c}{$\Gamma_{\alpha}$$^{b}$} & \multicolumn{1}{c}{$\Gamma_{\gamma}$$^{c}$} & \multicolumn{2}{c}{$\Gamma_{n}$$^{d}$} & \multicolumn{2}{c|}{$\Gamma$} & Int                                                                         \\   \hline
          11318         & 830.8 (13)                                       & $2^{+}$             & $1.4 (3) $                             & $10^{-4}$                                   & \multicolumn{2}{c}{- - - -}                 & - - - -                                & \multicolumn{2}{c}{- - - -}   &    2.5 (17)                  &   $10^2$           &                                \\    
          11441         & 976.39 (23)                                      & $4^{+}$             & $3.9 (10) $                               & $10^{-5}$                                   & $ 4.3 (11) $                                 & $ 10^{-6}$                             & 3.0 (15)                      & 1.47 (8)                     & $10^{3}$ & 1.47 (8)  & $10^{3}$ & \checkmark \\
          11465         & 1005.23 (25)                                     & $5^{-}$             & $5.5 (17) $                              & $10^{-5}$                                   & $ 5.0 (15) $                                & $ 10^{-6}$                             & 3.0 (15)                      & 6.55 (9)                     & $10^{3}$ & 6.55 (9)  & $10^{3}$ & \checkmark \\
          11508         & 1055.9 (11)                                      & $1^{-}$             & $3.5 (6) $                               & $10^{-4}$                                   & $ 1.17 (20) $                               & $ 10^{-4}$                             & 3.0 (15)                      & 1.27 (25)                    & $10^{4}$ & 1.27 (25) & $10^{4}$ & \checkmark \\
          11525         & 1075.5 (18)                                      & $1^{-}$             & $1.3 (3) $                               & $10^{-3}$                                   & $ 4.3 (11) $                                 & $ 10^{-4}$                             & 3.0 (15)                      & 1.8 (9)                      & $10^{3}$ & 1.8 (9)   & $10^{3}$ & \checkmark \\
          11632         & 1202.3 (17)                                      & $1^{-}$             & $7.1 (15) $                              & $10^{-3}$                                   & $ 2.4 (5) $                                 & $ 10^{-3}$                             & 3.0 (15)                      & 1.35 (17)                    & $10^{4}$ & 1.35 (17) & $10^{4}$ & \checkmark \\
          11752         & 1345 (7)                                         & $1^{-}$             & $5.9 (8) $                               & $10^{-2}$                                   & $ 2.0 (3) $                                 & $ 10^{-2}$                             & 3.0 (15)                      & 6.4 (9)                      & $10^{4}$ & 6.4 (9)   & $10^{4}$ & \checkmark \\
          11788         & 1386 (3)                                         & $1^{-}$             & $2.5 (9) $                               & $10^{-2}$                                   & $ 8 (2) $                                & $ 10^{-3}$                             & 3.0 (15)                      & 2.45 (24)                    & $10^{4}$ & 2.45 (24) & $10^{4}$ & \checkmark \\
          11828         & 1433.7 (12)                                      & $2^{+}$             & $8.5 (14) $                              & $10^{-1}$                                   & $ 1.7 (3)$                                  & $ 10^{-1}$                             & 3.0 (15)                      & 1.10 (25)                    & $10^{3}$ & 1.10 (25) & $10^{3}$ & \checkmark \\
          11863         & 1475 (3)                                         & $1^{-}$             & $5 (3) $                                 & $10^{-2}$                                   & $ 1.5 (10) $                                & $ 10^{-2}$                             & 3.0 (15)                      & 2.45 (34)                    & $10^{4}$ & 2.45 (34) & $10^{4}$ & \checkmark \\
          11880         & 1495 (3)                                         & $1^{-}$             & $1.9 (19) $                              & $10^{-1}$                                   & \multicolumn{2}{c}{- - - -}                 & - - - -                                & \multicolumn{2}{c}{- - - -}   & \multicolumn{2}{c|}{- - - -} &                                              \\
          11890         & 1507.9 (16)                                      & $1^{-}$             & $4.1 (4) $                               & $10^{-1}$                                   & \multicolumn{2}{c}{- - - -}                 & - - - -                                & \multicolumn{2}{c}{- - - -}   & \multicolumn{2}{c|}{- - - -} &                                              \\
          11910         & 1530.9 (15)                                      & $\mathbf{1^{-}},2^{+}$       & $1.40 (10) $                             & $10^{+0}$                                   & \multicolumn{2}{c}{- - - -}                 & - - - -                                & \multicolumn{2}{c}{- - - -}   & \multicolumn{2}{c|}{- - - -} &                                              \\
          11951         & 1579.4 (15)                                      & $2^{+},\mathbf{3^{-}},4^{+}$ & $1.60 (13) $                             & $10^{+0}$                                   & \multicolumn{2}{c}{- - - -}                 & - - - -                                & \multicolumn{2}{c}{- - - -}   & \multicolumn{2}{c|}{- - - -} &                                              \\    
          12050         & 1696.7 (15)                                      & $\mathbf{2^{+}},3^{-}$       & $4.7 (3) $                               & $10^{+0}$                                   & \multicolumn{2}{c}{- - - -}                 & - - - -                                & \multicolumn{2}{c}{- - - -}   & \multicolumn{2}{c|}{- - - -} &                                              \\
          12111         & 1768.2 (18)                                      & $1^{-}$             & $7.1 (6) $                               & $10^{-1}$                                   & \multicolumn{2}{c}{- - - -}                 & - - - -                                & \multicolumn{2}{c}{- - - -}   & \multicolumn{2}{c|}{- - - -} &                                              \\
          12141         & 1803.5 (15)                                      & $1^{-}$             & $2.4 (2) $                               & $10^{+0}$                                   & \multicolumn{2}{c}{- - - -}                 & - - - -                                & \multicolumn{2}{c}{- - - -}   & \multicolumn{2}{c|}{- - - -} &                                              \\
          12184         & 1855 (6)                                         & ($0^{+}$)           & $9.0 (11) $                              & $10^{-1}$                                   & \multicolumn{2}{c}{- - - -}                 & - - - -                                & \multicolumn{2}{c}{- - - -}   & \multicolumn{2}{c|}{- - - -} &                                              \\
          12270         & 1956 (6)                                         & ($0^{+}$)           & $2.1 (2) $                               & $10^{+1}$                                   & \multicolumn{2}{c}{- - - -}                 & - - - -                                & \multicolumn{2}{c}{- - - -}   & \multicolumn{2}{c|}{- - - -} &                                              \\
          12345         & 2044.8 (18)                                      & $0^{+}$             & $1.57 (10) $                             & $10^{+2}$                                   & \multicolumn{2}{c}{- - - -}                 & - - - -                                & \multicolumn{2}{c}{- - - -}   & \multicolumn{2}{c|}{- - - -} &                                              \\
          12435         & 2152 (10)                                        & $1^{-}$             & $2.8 (7) $                               & $10^{+1}$                                   & \multicolumn{2}{c}{- - - -}                 & - - - -                                & \multicolumn{2}{c}{- - - -}   & \multicolumn{2}{c|}{- - - -} &                                              \\
          12551         & 2289 (15)                                        & $1^{-}$             & $1.2 (5) $                               & $10^{+2}$                                   & \multicolumn{2}{c}{- - - -}                 & - - - -                                & \multicolumn{2}{c}{- - - -}   & \multicolumn{2}{c|}{- - - -} &                                              \\
          \hline \hline 
        \end{tabular} 
      } 
    \footnotetext{A detailed discussion of quantum number
      assignment can be found in the text.}
    \footnotetext{Calculated using equation~(\ref{eqn:approxGa}).}
    \footnotetext{Average value from Ref.\ \cite{KOE02}.}
    \footnotetext{Assuming $\Gamma$ is dominated by $\Gamma_{n}$
      (see section~\ref{sec:general-aspects}).}
    \caption{\label{tab:an_direct}Resonances in \an with known $\alpha$-particle
      partial widths or resonance strengths. When a range of quantum
      numbers is present, the one used for calculating the reaction
      rates is presented in bold.}
  \end{center} 
\end{table*}

\begin{table*}[!ht]
  \begin{center}
    { 

        \begin{tabular}{cccr@{$\times$}l|cr@{$\times$}lccc|c}
          \hline \hline 
          &                                                &           & \multicolumn{2}{c|}{ }                                 & \multicolumn{6}{c|}{Partial Widths (eV)}          &                                                                                                                                                           \\
          E$_{x}$ (keV) & \multicolumn{1}{c}{E$_{r}^{\textrm{lab}}$   (keV)} & J$^{\pi}$ & \multicolumn{2}{c|}{$\omega\gamma_{\mathrm{UL}}$ (eV)} & S$_{\alpha}$ & \multicolumn{2}{c}{$\Gamma_{\alpha,\mathrm{UL}}$} & \multicolumn{1}{c}{$\Gamma_{n}$} & \multicolumn{1}{c}{$\Gamma_{\gamma}$} & \multicolumn{1}{c|}{$\Gamma$} & Int                                        \\ \hline 
          11112         & 587.90 (10)        & $2^{+}$   & 5.8         & $10^{-8}$       &    1.00                  & 7.7                              & $10^{-9}$                             & $2580   (240)$                    & $1.73 (3)$   & $2580 (240) $  & \checkmark \\    
          11163         & 647.93 (11)        & $2^{+}$   & 1.9         & $10^{-7}$       &    0.44                  & 3.8                              & $10^{-8}$                             & $4640   (100) $                   & $4.56 (29) $ & $4650 (100)$   & \checkmark \\
          11171         & 657.53 (19)        & $(2^{+})$ & 7.5         & $10^{-8}$       &    0.12                  & 1.5                              & $10^{-8}$                             & $1.44   (16) $                    & $3.0 (15)$         & $4.4 (15) $    &            \\
          11183         & 671.70 (21)        & $(1^{-})$ & 7.7         & $10^{-5}$       &    1.00                  & 2.1                              & $10^{-7}$                             & $0.54 (9) $                       & $3.0 (15)  $       & $3.5 (15) $    &            \\
          11243         & 742.81 (12)        & $2^{(-)}$ & 1.2         & $10^{-7}$       &    0.01                  & 2.4                              & $10^{-8}$                             & $4510 (110) $                     & $7.4 (6) $   & $4520   (110)$ & \checkmark \\
          11274         & 779.33 (14)        & $(2)^{+}$ & 1.1         & $10^{-7}$       &    3.5$\times 10^{-3}$    & 2.2                              & $10^{-8}$                             & $540 (50) $                       & $3.2 (4) $   & $540   (50) $  & \checkmark \\
          11280         & 786.17 (13)        & $4^{(-)}$ & 1.3         & $10^{-7}$       &    0.16                   & 1.4                              & $10^{-8}$                             & $1510 (30) $                      & $0.59 (24) $ & $1510   (30) $ & \checkmark \\
          11286         & 792.90 (15)        & $1^{-}$   & 7.7         & $10^{-8}$       &    7.3$\times 10^{-4}$    & 2.6                              & $10^{-8}$                             & $1260 (100) $                     & $0.8 (5) $   & $1260   (100)$ & \checkmark \\
          11286         & 793.83 (14)        & $(2^{+})$ & 7.7         & $10^{-8}$       &    1.6$\times 10^{-3}$    & 1.5                              & $10^{-8}$                             & $13 (6) $                         & $4.3 (6) $   & $17 (6) $      & \checkmark \\
          11289         & 797.10 (29)        & $(2^{-})$ & 7.7         & $10^{-8}$       &    1.5$\times 10^{-3}$    & 1.5                              & $10^{-8}$                             & $1.5 (5) $                        & $3.0 (15)$         & $4.5   (16) $  &            \\
          11296         & 805.19 (16)        & $(3^{-})$ & 1.0         & $10^{-7}$       &    7.7$\times 10^{-3}$    & 1.4                              & $10^{-8}$                             & $8060 (120) $                     & $3.3 (7) $   & $8060 (120)$   & \checkmark \\
          11311         & 822.6 (4)          & $(1^{-})$ & 1.6         & $10^{-8}$       &    7.5$\times 10^{-5}$    & 5.8                              & $10^{-9}$                             & $1.1 (4) $                        & $3.0 (15)$         & $4.1 (16) $    &            \\
          11326         & 840.8 (6)          & $(1^{-})$ & 1.2         & $10^{-7}$       &    3.6$\times 10^{-4}$    & 4.5                              & $10^{-8}$                             & $0.6 (3) $                        & $3.0 (15)$         & $3.6 (15) $    &            \\
          11328         & 843.24 (17)        & $1^{-}$   & 5.0         & $10^{-7}$       &    1.3$\times 10^{-3}$    & 1.7                              & $10^{-7}$                             & $420   (90) $                     & $3.6 (5) $   & $430 (90) $    & \checkmark \\
          11329         & 844.4 (6)          & $(1^{-})$ & 1.2         & $10^{-7}$       &    3.3$\times 10^{-4}$    & 4.5                              & $10^{-8}$                             & $2.8   (10) $                     & $3.0 (15)$         & $5.8 (18) $    &            \\
          11337         & 853.6 (7)          & $(1^{-})$ & 1.3         & $10^{-7}$       &    2.7$\times 10^{-4}$    & 4.6                              & $10^{-8}$                             & $1.4 (6) $                        & $3.0 (15)$         & $4.4 (18) $    &            \\
          11344         & 861.86 (18)        & $(2^{+})$ & 2.0         & $10^{-7}$       &    7.2$\times 10^{-4}$    & 4.0                              & $10^{-8}$                             & $150 (40) $                       & $1.18 (27) $ & $150   (40) $  & \checkmark \\
          11345         & 862.91 (19)        & $4^{(-)}$ & 4.2         & $10^{-8}$       &    7.2$\times 10^{-4}$    & 5.1                              & $10^{-9}$                             & $4130 (190) $                     & $1.8 (4) $   & $4130 (190)$   & \checkmark \\
          11393         & 919.34 (19)        & $5^{(+)}$ & 3.7         & $10^{-8}$       &    2.4$\times 10^{-2}$    & 3.7                              & $10^{-9}$                             & $290 (19) $                       & $3.0 (15)$         & $293   (19) $  & \checkmark \\
          \hline \hline 
        \end{tabular} 
      } 
    \caption{\label{tab:an_upperlims}Properties of Unobserved
      Resonances in \an. For these resonances, only upper
      limits of the resonance strength and/or the $\alpha$-particle
      spectroscopic factor can be derived.  Quantum numbers,
      $\gamma$-ray and neutron partial widths are taken from the
      R-matrix fit of Ref.\ \cite{KOE02}. Resonance energies represent
      a weighted average of values adopted from Refs.\
      \cite{MOS76,WEI76,GLA86,KOE02}.}
  \end{center}
\end{table*}

The matching temperature, $T_{\text{match}}$, (see Sec.\
\ref{sec-2_4}) for both the \ag and \an reactions, beyond which the
rates are estimated by normalising Hauser-Feshbach predictions to
experimental rates, amounts to $T = 1.33$\ GK i.e., well above the
temperatures relevant for the s-process during He-burning ($T = 0.01 -
0.3$\ GK). 

\begin{figure*} 
  \begin{center}
    \includegraphics[width=0.9\textwidth]{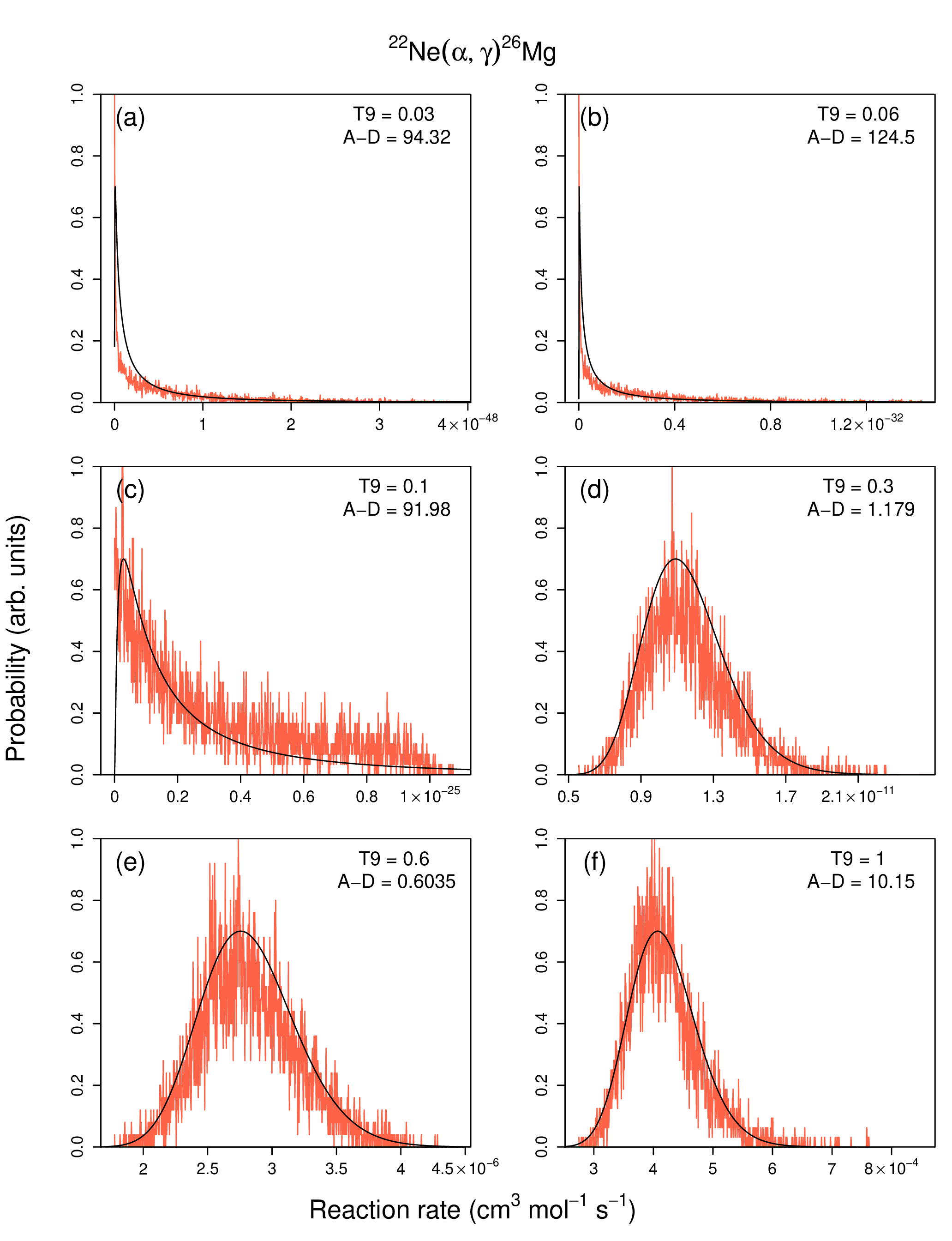} 
    \caption[\ag rate distributions]{\label{fig:Panel_ag}(Colour
      online) Reaction rate
      probability densities for the
      $^{22}$Ne($\alpha$,$\gamma$)$^{26}$Mg reaction at various
      stellar temperatures. 
      In each panel, the red histogram represents the Monte Carlo
      results, while the solid line shows the lognormal
      approximation. Note that the solid line is {\textit{not}} a fit
      to the histogram, but was calculated from the lognormal
      parameters $\mu$ and $\sigma$ (table~\ref{tab:agRate}), which in
      turn were determined from equation~(\ref{eq:rates-lognormPars}).
      It is apparent that the lognormal approximation to the reaction
      rates holds in the temperature range of the s-process (near 0.3\
      GK).}
  \end{center}
\end{figure*}

\begin{figure*} 
  \begin{center}
    \includegraphics[width=0.95\textwidth]{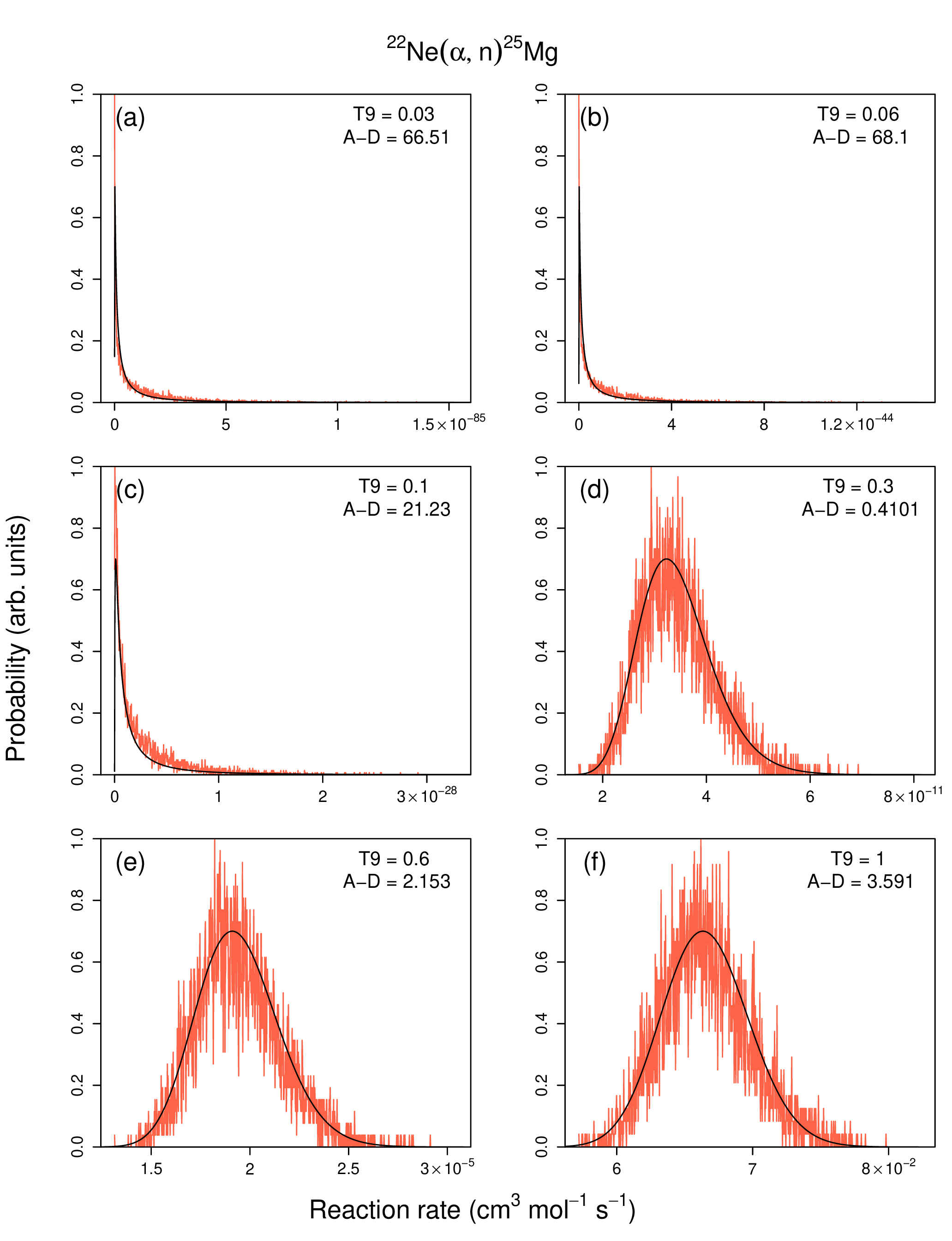} 
    \caption[\an rate distributions]{\label{fig:Panel_an}(Colour
      online) Reaction rate probability densities for the
      $^{22}$Ne($\alpha$,n)$^{25}$Mg reaction. See caption to
      Fig.~\ref{fig:Panel_ag}.}
  \end{center}
\end{figure*}

Monte Carlo reaction rates for the \ag and \an reactions are presented
in Tabs. \ref{tab:agRate} and\ \ref{tab:anRate}, respectively.  The
median, low, and high rates are shown alongside the lognormal
parameters and the Anderson-Darling statistic described in Sec.\
\ref{sec:rates-montecarlo}. 
The Monte Carlo reaction rate probability density functions are
displayed in Figs.~\ref{fig:Panel_ag} and~\ref{fig:Panel_an} as red
histograms. The solid black lines indicate the lognormal
approximation, calculated with the lognormal parameters, $\mu$ and
$\sigma$, listed in columns 5, 6, 10, and 11 of Tabs.~\ref{tab:agRate}
and~\ref{tab:anRate}.

\begin{table*} 
  \begin{center} 
      \begin{tabular}{l||ccc|cc} 
        \hline \hline 
        T (GK) & Low rate & Median rate & High rate & lognormal $\mu$ & lognormal $\sigma$ \\ \hline 
        0.010 &  1.05$\times$10$^{-77}$  &  2.14$\times$10$^{-77}$  &        4.52$\times$10$^{-77}$  &  -1.765$\times$10$^{+02}$  &        7.42$\times$10$^{-01}$  \\
        0.011 &  3.99$\times$10$^{-74}$  &  7.28$\times$10$^{-74}$  &        1.34$\times$10$^{-73}$  &  -1.684$\times$10$^{+02}$  &        6.15$\times$10$^{-01}$  \\
        0.012 &  3.69$\times$10$^{-71}$  &  6.34$\times$10$^{-71}$  &        1.07$\times$10$^{-70}$  &  -1.617$\times$10$^{+02}$  &        5.34$\times$10$^{-01}$  \\
        0.013 &  1.15$\times$10$^{-68}$  &  1.90$\times$10$^{-68}$  &        3.09$\times$10$^{-68}$  &  -1.559$\times$10$^{+02}$  &        4.92$\times$10$^{-01}$  \\
        0.014 &  1.55$\times$10$^{-66}$  &  2.52$\times$10$^{-66}$  &        4.04$\times$10$^{-66}$  &  -1.511$\times$10$^{+02}$  &        4.80$\times$10$^{-01}$  \\
        0.015 &  1.06$\times$10$^{-64}$  &  1.73$\times$10$^{-64}$  &        2.79$\times$10$^{-64}$  &  -1.468$\times$10$^{+02}$  &        4.90$\times$10$^{-01}$  \\
        0.016 &  4.11$\times$10$^{-63}$  &  6.96$\times$10$^{-63}$  &        1.14$\times$10$^{-62}$  &  -1.431$\times$10$^{+02}$  &        5.13$\times$10$^{-01}$  \\
        0.018 &  1.80$\times$10$^{-60}$  &  3.26$\times$10$^{-60}$  &        5.63$\times$10$^{-60}$  &  -1.370$\times$10$^{+02}$  &        5.75$\times$10$^{-01}$  \\
        0.020 &  2.24$\times$10$^{-58}$  &  4.34$\times$10$^{-58}$  &        8.04$\times$10$^{-58}$  &  -1.321$\times$10$^{+02}$  &        6.43$\times$10$^{-01}$  \\
        0.025 &  1.54$\times$10$^{-54}$  &  3.14$\times$10$^{-54}$  &        6.30$\times$10$^{-54}$  &  -1.232$\times$10$^{+02}$  &        7.13$\times$10$^{-01}$  \\
        0.030 &  2.82$\times$10$^{-50}$  &  3.35$\times$10$^{-49}$  &        1.30$\times$10$^{-48}$  &  -1.121$\times$10$^{+02}$  &        1.87$\times$10$^{+00}$  \\
        0.040 &  1.81$\times$10$^{-42}$  &  2.31$\times$10$^{-41}$  &        8.91$\times$10$^{-41}$  &  -9.413$\times$10$^{+01}$  &        2.14$\times$10$^{+00}$  \\
        0.050 &  8.51$\times$10$^{-38}$  &  1.08$\times$10$^{-36}$  &        4.17$\times$10$^{-36}$  &  -8.338$\times$10$^{+01}$  &        2.15$\times$10$^{+00}$  \\
        0.060 &  1.05$\times$10$^{-34}$  &  1.34$\times$10$^{-33}$  &        5.14$\times$10$^{-33}$  &  -7.624$\times$10$^{+01}$  &        2.08$\times$10$^{+00}$  \\
        0.070 &  1.95$\times$10$^{-32}$  &  2.12$\times$10$^{-31}$  &        8.04$\times$10$^{-31}$  &  -7.104$\times$10$^{+01}$  &        1.79$\times$10$^{+00}$  \\
        0.080 &  2.76$\times$10$^{-30}$  &  1.14$\times$10$^{-29}$  &        3.67$\times$10$^{-29}$  &  -6.679$\times$10$^{+01}$  &        1.33$\times$10$^{+00}$  \\
        0.090 &  1.76$\times$10$^{-28}$  &  6.30$\times$10$^{-28}$  &        1.35$\times$10$^{-27}$  &  -6.289$\times$10$^{+01}$  &        1.15$\times$10$^{+00}$  \\
        0.100 &  4.79$\times$10$^{-27}$  &  2.28$\times$10$^{-26}$  &        6.55$\times$10$^{-26}$  &  -5.931$\times$10$^{+01}$  &        1.35$\times$10$^{+00}$  \\
        0.110 &  8.17$\times$10$^{-26}$  &  5.95$\times$10$^{-25}$  &        1.86$\times$10$^{-24}$  &  -5.616$\times$10$^{+01}$  &        1.55$\times$10$^{+00}$  \\
        0.120 &  1.11$\times$10$^{-24}$  &  9.63$\times$10$^{-24}$  &        3.07$\times$10$^{-23}$  &  -5.343$\times$10$^{+01}$  &        1.64$\times$10$^{+00}$  \\
        0.130 &  1.23$\times$10$^{-23}$  &  1.03$\times$10$^{-22}$  &        3.28$\times$10$^{-22}$  &  -5.102$\times$10$^{+01}$  &        1.57$\times$10$^{+00}$  \\
        0.140 &  1.38$\times$10$^{-22}$  &  8.23$\times$10$^{-22}$  &        2.50$\times$10$^{-21}$  &  -4.883$\times$10$^{+01}$  &        1.36$\times$10$^{+00}$  \\
        0.150 &  1.53$\times$10$^{-21}$  &  5.57$\times$10$^{-21}$  &        1.51$\times$10$^{-20}$  &  -4.679$\times$10$^{+01}$  &        1.10$\times$10$^{+00}$  \\
        0.160 &  1.41$\times$10$^{-20}$  &  3.79$\times$10$^{-20}$  &        8.10$\times$10$^{-20}$  &  -4.484$\times$10$^{+01}$  &        8.63$\times$10$^{-01}$  \\
        0.180 &  8.05$\times$10$^{-19}$  &  1.54$\times$10$^{-18}$  &        2.84$\times$10$^{-18}$  &  -4.102$\times$10$^{+01}$  &        6.29$\times$10$^{-01}$  \\
        0.200 &  3.41$\times$10$^{-17}$  &  5.43$\times$10$^{-17}$  &        9.60$\times$10$^{-17}$  &  -3.740$\times$10$^{+01}$  &        5.19$\times$10$^{-01}$  \\
        0.250 &  5.88$\times$10$^{-14}$  &  7.56$\times$10$^{-14}$  &        1.00$\times$10$^{-13}$  &  -3.019$\times$10$^{+01}$  &        2.78$\times$10$^{-01}$  \\
        0.300 &  9.32$\times$10$^{-12}$  &  1.13$\times$10$^{-11}$  &        1.38$\times$10$^{-11}$  &  -2.520$\times$10$^{+01}$  &        1.96$\times$10$^{-01}$  \\
        0.350 &  3.46$\times$10$^{-10}$  &  4.08$\times$10$^{-10}$  &        4.86$\times$10$^{-10}$  &  -2.162$\times$10$^{+01}$  &        1.69$\times$10$^{-01}$  \\
        0.400 &  5.11$\times$10$^{-09}$  &  5.95$\times$10$^{-09}$  &        6.98$\times$10$^{-09}$  &  -1.894$\times$10$^{+01}$  &        1.56$\times$10$^{-01}$  \\
        0.450 &  4.09$\times$10$^{-08}$  &  4.72$\times$10$^{-08}$  &        5.50$\times$10$^{-08}$  &  -1.686$\times$10$^{+01}$  &        1.47$\times$10$^{-01}$  \\
        0.500 &  2.13$\times$10$^{-07}$  &  2.44$\times$10$^{-07}$  &        2.82$\times$10$^{-07}$  &  -1.522$\times$10$^{+01}$  &        1.41$\times$10$^{-01}$  \\
        0.600 &  2.47$\times$10$^{-06}$  &  2.79$\times$10$^{-06}$  &        3.20$\times$10$^{-06}$  &  -1.278$\times$10$^{+01}$  &        1.32$\times$10$^{-01}$  \\
        0.700 &  1.39$\times$10$^{-05}$  &  1.57$\times$10$^{-05}$  &        1.78$\times$10$^{-05}$  &  -1.106$\times$10$^{+01}$  &        1.25$\times$10$^{-01}$  \\
        0.800 &  5.15$\times$10$^{-05}$  &  5.77$\times$10$^{-05}$  &        6.51$\times$10$^{-05}$  &  -9.758$\times$10$^{+00}$  &        1.18$\times$10$^{-01}$  \\
        0.900 &  1.48$\times$10$^{-04}$  &  1.66$\times$10$^{-04}$  &        1.88$\times$10$^{-04}$  &  -8.701$\times$10$^{+00}$  &        1.19$\times$10$^{-01}$  \\
        1.000 &  3.65$\times$10$^{-04}$  &  4.11$\times$10$^{-04}$  &        4.73$\times$10$^{-04}$  &  -7.788$\times$10$^{+00}$  &        1.35$\times$10$^{-01}$  \\
        1.250 &  2.33$\times$10$^{-03}$  &  2.77$\times$10$^{-03}$  &        3.43$\times$10$^{-03}$  &  -5.867$\times$10$^{+00}$  &        2.02$\times$10$^{-01}$  \\
        1.500 & (1.45$\times$10$^{-02}$) & (1.79$\times$10$^{-02}$) &        (2.21$\times$10$^{-02}$) & (-4.024$\times$10$^{+00}$) &      (2.12$\times$10$^{-01}$) \\
        1.750 & (7.64$\times$10$^{-02}$) & (9.45$\times$10$^{-02}$) &        (1.17$\times$10$^{-01}$) & (-2.360$\times$10$^{+00}$) &      (2.12$\times$10$^{-01}$) \\
        2.000 & (3.00$\times$10$^{-01}$) & (3.70$\times$10$^{-01}$) &        (4.58$\times$10$^{-01}$) & (-9.932$\times$10$^{-01}$) &      (2.12$\times$10$^{-01}$) \\
        2.500 & (2.55$\times$10$^{+00}$) & (3.15$\times$10$^{+00}$) &        (3.89$\times$10$^{+00}$) & (1.147$\times$10$^{+00}$) &       (2.12$\times$10$^{-01}$) \\
        3.000 & (1.24$\times$10$^{+01}$) & (1.53$\times$10$^{+01}$) &        (1.89$\times$10$^{+01}$) & (2.729$\times$10$^{+00}$) &       (2.12$\times$10$^{-01}$) \\
        3.500 & (4.18$\times$10$^{+01}$) & (5.17$\times$10$^{+01}$) &        (6.39$\times$10$^{+01}$) & (3.945$\times$10$^{+00}$) &       (2.12$\times$10$^{-01}$) \\
        4.000 & (1.10$\times$10$^{+02}$) & (1.36$\times$10$^{+02}$) &        (1.68$\times$10$^{+02}$) & (4.913$\times$10$^{+00}$) &       (2.12$\times$10$^{-01}$) \\
        5.000 & (4.71$\times$10$^{+02}$) & (5.82$\times$10$^{+02}$) &        (7.19$\times$10$^{+02}$) & (6.366$\times$10$^{+00}$) &       (2.12$\times$10$^{-01}$) \\
        6.000 & (1.33$\times$10$^{+03}$) & (1.64$\times$10$^{+03}$) &        (2.03$\times$10$^{+03}$) & (7.405$\times$10$^{+00}$) &       (2.12$\times$10$^{-01}$) \\
        7.000 & (2.91$\times$10$^{+03}$) & (3.59$\times$10$^{+03}$) &        (4.44$\times$10$^{+03}$) & (8.186$\times$10$^{+00}$) &       (2.12$\times$10$^{-01}$) \\
        8.000 & (5.35$\times$10$^{+03}$) & (6.62$\times$10$^{+03}$) &        (8.18$\times$10$^{+03}$) & (8.798$\times$10$^{+00}$) &       (2.12$\times$10$^{-01}$) \\
        9.000 & (8.68$\times$10$^{+03}$) & (1.07$\times$10$^{+04}$) &        (1.33$\times$10$^{+04}$) & (9.281$\times$10$^{+00}$) &       (2.12$\times$10$^{-01}$) \\
        10.000& (1.30$\times$10$^{+04}$) & (1.60$\times$10$^{+04}$) &        (1.98$\times$10$^{+04}$) & (9.681$\times$10$^{+00}$) &      (2.12$\times$10$^{-01}$) \\
        \hline   \hline 
      \end{tabular}
      \caption{\label{tab:agRate} Monte Carlo reaction rates for the
      $^{22}$Ne($\alpha$,$\gamma$)$^{26}$Mg reaction. Shown are the
      low, median, and high rates, corresponding to the 16th, 50th,
      and 84th percentiles of the Monte Carlo probability density
      distributions. Also shown are the parameters ($\mu$ and
      $\sigma$) of the lognormal approximation to the actual Monte
      Carlo probability density. See Ref.\ \cite{LON10} for
      details. The rate values shown in parentheses indicate the
      temperatures ($T>T_{\text{match}}=1.33$~GK) for which
      Hauser-Feshbach rates, normalised to experimental results, are
      adopted (see section~\ref{sec-2_4}).}
  \end{center} 
\end{table*}

\begin{table*} 
  \begin{center} 
      \begin{tabular}{l||ccc|cc} 
        \hline \hline 
        T (GK) & Low rate & Median rate & High rate & lognormal $\mu$ & lognormal $\sigma$ \\ \hline 
        0.010 &  0.0  &  0.0  &          0.0  &  - - - - &          - - - -    \\ 
        0.011 &  0.0  &  0.0  &          0.0  &  - - - - &          - - - -    \\ 
        0.012 &  0.0  &  0.0  &          0.0  &  - - - - &          - - - -    \\ 
        0.013 &  0.0  &  0.0  &          0.0  &  - - - - &          - - - -    \\ 
        0.014 &  0.0  &  0.0  &          0.0  &  - - - - &          - - - -    \\ 
        0.015 &  0.0  &  0.0  &          0.0  &  - - - - &          - - - -    \\ 
        0.016 &  0.0  &  0.0  &          0.0  &  - - - - &          - - - -    \\ 
        0.018 &  0.0  &  0.0  &          0.0  &  - - - - &          - - - -    \\ 
        0.020 &  0.0  &  0.0  &          0.0  &  - - - - &          - - - -    \\ 
        0.025 &  0.0  &  0.0  &          0.0  &  - - - - &          - - - -    \\ 
        0.030 &  5.12$\times$10$^{-88}$   &  5.08$\times$10$^{-87}$   &          2.25$\times$10$^{-86}$   &  -1.991$\times$10$^{+02}$ &          1.90$\times$10$^{+00}$    \\ 
        0.040 &  1.46$\times$10$^{-67}$   &  1.49$\times$10$^{-66}$   &          6.64$\times$10$^{-66}$   &  -1.519$\times$10$^{+02}$ &          1.94$\times$10$^{+00}$    \\ 
        0.050 &  2.99$\times$10$^{-55}$   &  3.05$\times$10$^{-54}$   &          1.36$\times$10$^{-53}$   &  -1.236$\times$10$^{+02}$ &          1.95$\times$10$^{+00}$    \\ 
        0.060 &  4.92$\times$10$^{-47}$   &  4.87$\times$10$^{-46}$   &          2.17$\times$10$^{-45}$   &  -1.047$\times$10$^{+02}$ &          1.92$\times$10$^{+00}$    \\ 
        0.070 &  3.70$\times$10$^{-41}$   &  3.48$\times$10$^{-40}$   &          1.55$\times$10$^{-39}$   &  -9.117$\times$10$^{+01}$ &          1.84$\times$10$^{+00}$    \\ 
        0.080 &  1.03$\times$10$^{-36}$   &  8.44$\times$10$^{-36}$   &          3.73$\times$10$^{-35}$   &  -8.101$\times$10$^{+01}$ &          1.74$\times$10$^{+00}$    \\ 
        0.090 &  3.23$\times$10$^{-33}$   &  2.19$\times$10$^{-32}$   &          9.43$\times$10$^{-32}$   &  -7.309$\times$10$^{+01}$ &          1.62$\times$10$^{+00}$    \\ 
        0.100 &  2.17$\times$10$^{-30}$   &  1.20$\times$10$^{-29}$   &          4.92$\times$10$^{-29}$   &  -6.673$\times$10$^{+01}$ &          1.50$\times$10$^{+00}$    \\ 
        0.110 &  4.65$\times$10$^{-28}$   &  2.12$\times$10$^{-27}$   &          8.22$\times$10$^{-27}$   &  -6.151$\times$10$^{+01}$ &          1.39$\times$10$^{+00}$    \\ 
        0.120 &  4.24$\times$10$^{-26}$   &  1.62$\times$10$^{-25}$   &          5.82$\times$10$^{-25}$   &  -5.714$\times$10$^{+01}$ &          1.29$\times$10$^{+00}$    \\ 
        0.130 &  1.94$\times$10$^{-24}$   &  6.61$\times$10$^{-24}$   &          2.14$\times$10$^{-23}$   &  -5.342$\times$10$^{+01}$ &          1.19$\times$10$^{+00}$    \\ 
        0.140 &  5.27$\times$10$^{-23}$   &  1.64$\times$10$^{-22}$   &          4.81$\times$10$^{-22}$   &  -5.020$\times$10$^{+01}$ &          1.08$\times$10$^{+00}$    \\ 
        0.150 &  9.94$\times$10$^{-22}$   &  2.74$\times$10$^{-21}$   &          7.18$\times$10$^{-21}$   &  -4.737$\times$10$^{+01}$ &          9.62$\times$10$^{-01}$    \\ 
        0.160 &  1.43$\times$10$^{-20}$   &  3.39$\times$10$^{-20}$   &          7.89$\times$10$^{-20}$   &  -4.484$\times$10$^{+01}$ &          8.29$\times$10$^{-01}$    \\ 
        0.180 &  1.61$\times$10$^{-18}$   &  2.74$\times$10$^{-18}$   &          5.01$\times$10$^{-18}$   &  -4.040$\times$10$^{+01}$ &          5.53$\times$10$^{-01}$    \\ 
        0.200 &  9.14$\times$10$^{-17}$   &  1.24$\times$10$^{-16}$   &          1.79$\times$10$^{-16}$   &  -3.660$\times$10$^{+01}$ &          3.43$\times$10$^{-01}$    \\ 
        0.250 &  1.68$\times$10$^{-13}$   &  2.06$\times$10$^{-13}$   &          2.53$\times$10$^{-13}$   &  -2.921$\times$10$^{+01}$ &          2.06$\times$10$^{-01}$    \\ 
        0.300 &  2.74$\times$10$^{-11}$   &  3.36$\times$10$^{-11}$   &          4.15$\times$10$^{-11}$   &  -2.411$\times$10$^{+01}$ &          2.06$\times$10$^{-01}$    \\ 
        0.350 &  1.05$\times$10$^{-09}$   &  1.29$\times$10$^{-09}$   &          1.59$\times$10$^{-09}$   &  -2.046$\times$10$^{+01}$ &          2.05$\times$10$^{-01}$    \\ 
        0.400 &  1.64$\times$10$^{-08}$   &  2.00$\times$10$^{-08}$   &          2.45$\times$10$^{-08}$   &  -1.773$\times$10$^{+01}$ &          1.99$\times$10$^{-01}$    \\ 
        0.450 &  1.42$\times$10$^{-07}$   &  1.71$\times$10$^{-07}$   &          2.07$\times$10$^{-07}$   &  -1.558$\times$10$^{+01}$ &          1.88$\times$10$^{-01}$    \\ 
        0.500 &  8.51$\times$10$^{-07}$   &  1.00$\times$10$^{-06}$   &          1.19$\times$10$^{-06}$   &  -1.381$\times$10$^{+01}$ &          1.68$\times$10$^{-01}$    \\ 
        0.600 &  1.74$\times$10$^{-05}$   &  1.92$\times$10$^{-05}$   &          2.15$\times$10$^{-05}$   &  -1.085$\times$10$^{+01}$ &          1.07$\times$10$^{-01}$    \\ 
        0.700 &  2.36$\times$10$^{-04}$   &  2.51$\times$10$^{-04}$   &          2.69$\times$10$^{-04}$   &  -8.287$\times$10$^{+00}$ &          6.70$\times$10$^{-02}$    \\ 
        0.800 &  2.15$\times$10$^{-03}$   &  2.27$\times$10$^{-03}$   &          2.42$\times$10$^{-03}$   &  -6.084$\times$10$^{+00}$ &          5.79$\times$10$^{-02}$    \\ 
        0.900 &  1.36$\times$10$^{-02}$   &  1.43$\times$10$^{-02}$   &          1.51$\times$10$^{-02}$   &  -4.246$\times$10$^{+00}$ &          5.33$\times$10$^{-02}$    \\ 
        1.000 &  6.34$\times$10$^{-02}$   &  6.64$\times$10$^{-02}$   &          6.98$\times$10$^{-02}$   &  -2.711$\times$10$^{+00}$ &          4.82$\times$10$^{-02}$    \\ 
        1.250 &  1.18$\times$10$^{+00}$   &  1.22$\times$10$^{+00}$   &          1.27$\times$10$^{+00}$   &  1.998$\times$10$^{-01}$  &          3.88$\times$10$^{-02}$    \\ 
        1.500 & (1.09$\times$10$^{+01}$)  & (1.14$\times$10$^{+01}$)  &          (1.18$\times$10$^{+01}$) & (2.431$\times$10$^{+00}$) &          (3.89$\times$10$^{-02}$)   \\
        1.750 & (6.79$\times$10$^{+01}$)  & (7.06$\times$10$^{+01}$)  &          (7.34$\times$10$^{+01}$) & (4.257$\times$10$^{+00}$) &          (3.89$\times$10$^{-02}$)   \\
        2.000 & (2.92$\times$10$^{+02}$)  & (3.04$\times$10$^{+02}$)  &          (3.16$\times$10$^{+02}$) & (5.717$\times$10$^{+00}$) &          (3.89$\times$10$^{-02}$)   \\
        2.500 & (2.74$\times$10$^{+03}$)  & (2.85$\times$10$^{+03}$)  &          (2.96$\times$10$^{+03}$) & (7.953$\times$10$^{+00}$) &          (3.89$\times$10$^{-02}$)  \\ 
        3.000 & (1.41$\times$10$^{+04}$)  & (1.46$\times$10$^{+04}$)  &          (1.52$\times$10$^{+04}$) & (9.590$\times$10$^{+00}$) &          (3.89$\times$10$^{-02}$)  \\ 
        3.500 & (4.96$\times$10$^{+04}$)  & (5.16$\times$10$^{+04}$)  &          (5.37$\times$10$^{+04}$) & (1.085$\times$10$^{+01}$) &          (3.89$\times$10$^{-02}$)  \\ 
        4.000 & (1.36$\times$10$^{+05}$)  & (1.41$\times$10$^{+05}$)  &          (1.47$\times$10$^{+05}$) & (1.186$\times$10$^{+01}$) &          (3.89$\times$10$^{-02}$)  \\ 
        5.000 & (6.10$\times$10$^{+05}$)  & (6.34$\times$10$^{+05}$)  &          (6.59$\times$10$^{+05}$) & (1.336$\times$10$^{+01}$) &          (3.89$\times$10$^{-02}$)  \\ 
        6.000 & (1.80$\times$10$^{+06}$)  & (1.88$\times$10$^{+06}$)  &          (1.95$\times$10$^{+06}$) & (1.444$\times$10$^{+01}$) &          (3.89$\times$10$^{-02}$)  \\ 
        7.000 & (4.07$\times$10$^{+06}$)  & (4.23$\times$10$^{+06}$)  &          (4.40$\times$10$^{+06}$) & (1.526$\times$10$^{+01}$) &          (3.89$\times$10$^{-02}$)  \\ 
        8.000 & (7.70$\times$10$^{+06}$)  & (8.01$\times$10$^{+06}$)  &          (8.32$\times$10$^{+06}$) & (1.590$\times$10$^{+01}$) &          (3.89$\times$10$^{-02}$)  \\ 
        9.000 & (1.28$\times$10$^{+07}$)  & (1.33$\times$10$^{+07}$)  &          (1.39$\times$10$^{+07}$) & (1.640$\times$10$^{+01}$) &          (3.89$\times$10$^{-02}$)  \\ 
        10.000&(1.97$\times$10$^{+07}$)   & (2.04$\times$10$^{+07}$)  &          (2.12$\times$10$^{+07}$) & (1.683$\times$10$^{+01}$) &          (3.89$\times$10$^{-02}$)   \\
        \hline   \hline 
      \end{tabular}
      \caption{\label{tab:anRate} Monte Carlo reaction rates for the
      $^{22}$Ne($\alpha$,n)$^{25}$Mg reaction. Shown are the low,
      median, and high rates, corresponding to the 16th, 50th, and
      84th percentiles of the Monte Carlo probability density
      distributions. Also shown are the parameters ($\mu$ and
      $\sigma$) of the lognormal approximation to the actual Monte
      Carlo probability density. See Ref.\ \cite{LON10} for
      details. The rate values shown in parentheses indicate the
      temperatures ($T>T_{\text{match}}=1.33$~GK) for which
      Hauser-Feshbach rates, normalised to experimental results, are
      adopted (see section~\ref{sec-2_4}).}
  \end{center} 
\end{table*}

\begin{figure*} 
  \begin{center}
    \includegraphics[width=0.7\textwidth]{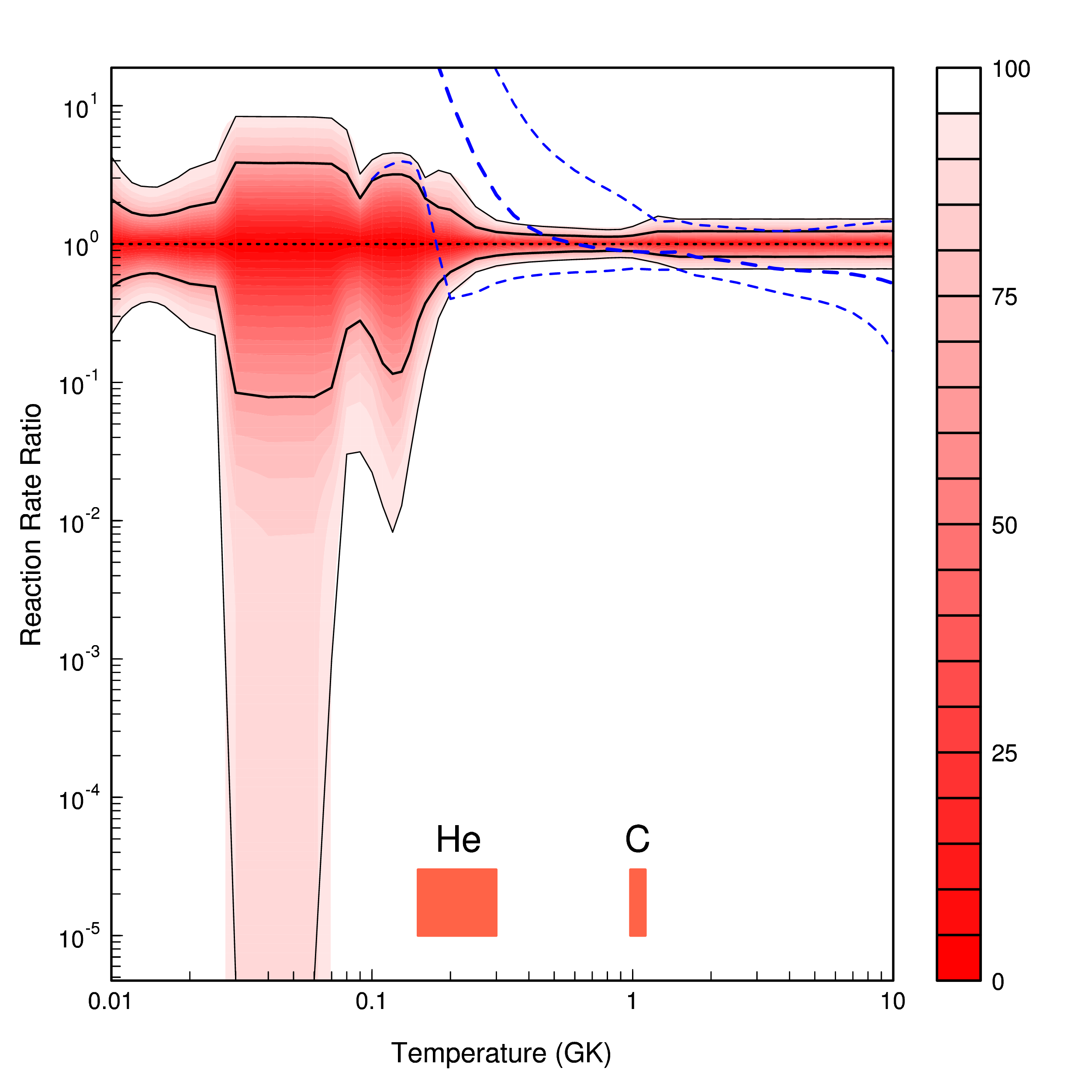}
    \caption[\ag comparison with NACRE]{\label{fig:Uncerts_ag}(Colour
      online) The uncertainty bands for the
      $^{22}$Ne($\alpha$,$\gamma$)$^{26}$Mg reaction. The
      uncertainties are the result of upper limit resonance
      contributions and of resonance strength uncertainties. The
      colour-densities represent the present reaction rate probability
      densities normalised to our recommended rate. The thick and thin
      black lines represent the 68\% and 95\% uncertainties,
      respectively. The dashed blue lines represent the literature
      rates from Ref.\ \cite{ANG99}, with the thick and thin lines
      denoting the recommended rate and rate limits, respectively,
      normalised to our recommended rate.  Values below unity (dotted
      line) indicate that the rates are lower than the present
      recommended rate. The relevant temperatures for helium- and
      carbon shell-burning have been added as red bars with the labels
      ``He'' and ``C'', respectively.}
  \end{center}
\end{figure*}

\begin{figure*} 
  \begin{center}
    \includegraphics[width=0.8\textwidth]{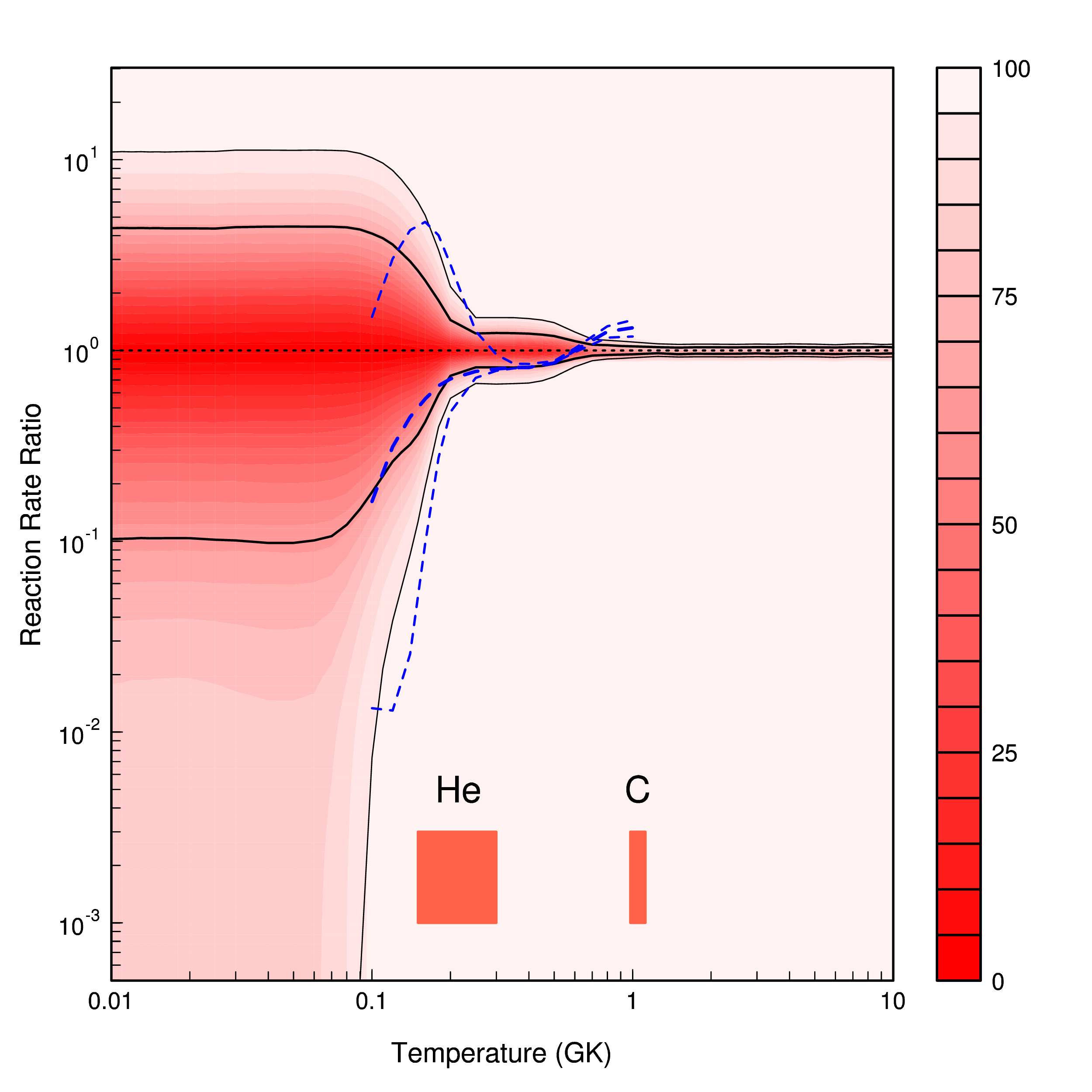}
    \caption[\an comparison with NACRE]{\label{fig:Uncerts_an}(Colour
      online) The probability densities for the
      $^{22}$Ne($\alpha$,n)$^{25}$Mg reaction in comparison to those
      presented by Ref.\ \cite{JAE01}. See caption of Fig.\
      \ref{fig:Uncerts_ag} for an explanation.}
  \end{center} 
\end{figure*}

In order to emphasise that our low and high rates, obtained for a
coverage probability of 68\% (see section~\ref{sec:rates-montecarlo}),
do {\textit{not}} represent sharp boundaries, we show the
($\alpha$,$\gamma$) and ($\alpha$,n) reaction rates, normalised to the
respective recommended (median) values, as colour contours in
Figs.~\ref{fig:Uncerts_ag} and~\ref{fig:Uncerts_an}. The thick
and thin solid lines represent coverage probabilities of 68\% and
95\%, respectively. The three dashed lines show the previously
reported rates (\textcite{ANG99} for
\reaction{22}{Ne}{$\alpha$}{$\gamma$}{26}{Mg} and \textcite{JAE01} for
\reaction{22}{Ne}{$\alpha$}{n}{25}{Mg}), normalised to our recommended
rate.  Our calculations of the relative resonance contributions to the
total ($\alpha$,$\gamma$) and ($\alpha$,n) reaction rates show that,
at temperatures most relevant to the s-process, resonances including
and below the \Erlab{831} resonance are the most important. Future
experimental efforts should, therefore, be concentrated on studying
resonances in the excitation energy region near the neutron threshold.

For the \ag reaction, the present rates deviate significantly from the
results of Ref.\ \cite{ANG99}, by factors of 2-100.
The differences are caused by: (i) a different treatment of partial
widths; in Ref.\ \cite{ANG99} the rates were found from numerical
integration by assuming upper limit values ($\Gamma=4-10$~keV) for the
total widths, whereas in the present work total widths have been
adopted from measured values; (ii) our improved treatment of upper
limits for reduced $\alpha$-particle widths (i.e., sampling over a
Porter-Thomas distribution; see section~\ref{sec:rates-montecarlo});
and (iii) the fact that new nuclear data became available since 1999
(see Tab.~\ref{tab:data-since-NACRE}). The combined effect of these
improvements results in a factor of 5 reduction in reaction rate
uncertainties in the He-burning temperature region.

As already noted in section~\ref{sec:general-aspects}, a number of
excited states near the $\alpha$-particle and neutron thresholds in
\nuc{26}{Mg} have been observed by additional inelastic proton
scattering experiments~\citep{MOS76,CRA89,TAM03}. However, 
their spins and parities have not been determined. In particular, it
is not known at present if these levels possess natural parity and
thereby may be populated in the \Nepa reactions.  In order to
investigate the maximum impact of these states with unknown $J^{\pi}$
values on the \ag reaction rate, we 
performed a test by assuming that all of these levels possess natural
parity and by adopting upper limit $\alpha$-particle spectroscopic
factors from that data of Refs.\ \cite{GIE93} and \cite{UGA07}. The
results show that these states can increase the \ag reaction rate by
up to a factor of 30 at temperatures between T$_9=0.1$ and $0.2$~GK.

For the \an reaction there is better agreement between previous and
new rates. The present rates are slightly higher (up to a factor of 2)
than those calculated by Ref.\ \cite{JAE01}. The two main reasons for
the difference are: (i) we used inflated weighted averages of the
reported resonance strengths from different measurements (see
section~\ref{sec:resonance-strengths}); and (ii) excluded the
contribution of a presumed \Erlab{630} resonance, because the level at
\Ex{11154} has been shown to possess unnatural parity~\citep{LON09}.

We would like to emphasise that the observed ($\alpha$,n) and
($\alpha$,$\gamma$) resonances near \Erlab{830} introduce another
systematic uncertainty that we have not accounted for. Recall that we
treated these two resonances as independent and narrow
(section~\ref{sec:levels}). On the other hand, if they correspond to
the same level in \nuc{26}{Mg}, the partial widths could be derived
from the measured resonance strengths. In that case, the resonance
turns out to be relatively broad, resulting in a significant
contribution of the resonance tail to the total reaction rate. Tests
show that the resulting reaction rates near $T \approx 0.3$~GK could
increase by roughly a factor of 5.

\begin{figure*}
  \centering
  \includegraphics[width=0.7\textwidth]{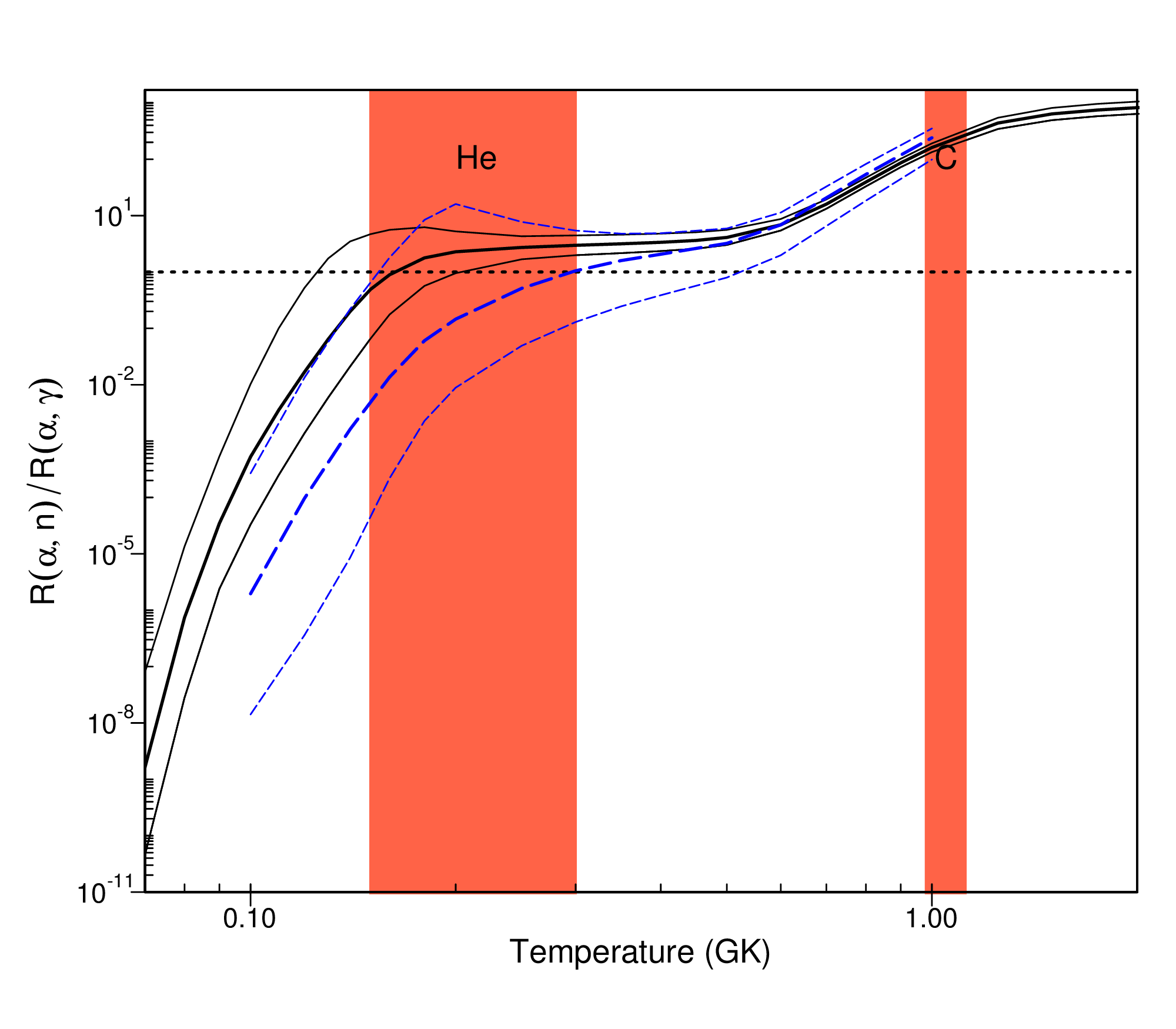}
  \caption{(Colour online) Uncertainty bands of the reaction rate
    \textit{ratio}, $N_{A}\langle \sigma v
    \rangle_{(\alpha,n)}/N_{A}\langle \sigma v
    \rangle_{(\alpha,\gamma)}$. The solid (black) lines represent
    the present reaction rate ratio, while the dashed (blue) lines
    represent the ratio of rates from Ref.\ \cite{JAE01} for
    \reaction{22}{Ne}{$\alpha$}{n}{25}{Mg}, and Ref.\ \cite{ANG99},
    for \reaction{22}{Ne}{$\alpha$}{$\gamma$}{26}{Mg}. The
    recommended ratio (the center line in each set) was calculated
    by dividing the recommended \an reaction rate by that of the
    \ag reaction at each temperature. To obtain the uncertainty
    bands for the rate ratio, the high rate for \ag was divided by
    the low rate for \reaction{22}{Ne}{$\alpha$}{n}{25}{Mg}, and
    vice versa. Values greater than unity indicate that more
    neutrons (and \nuc{25}{Mg}) are produced than $\gamma$-rays
    (and \nuc{26}{Mg}) per $\alpha$-particle capture. The
    temperatures relevant in helium- and carbon shell-burning are
    represented by red bars and are marked with ``He'' and ``C'',
    respectively.}
  \label{fig:RateRatio}
\end{figure*}

The ratio of \an to \ag reaction rates is shown in
figure~\ref{fig:RateRatio}. Note that these rates are not independent
since, for example, the same values of $\alpha$-particle partial
widths enter in both rate calculations if an $(\alpha,\gamma)$ and
$(\alpha,n)$ resonance corresponds to the same \nuc{26}{Mg}
level. Thus the uncertainties shown in Fig.~\ref{fig:RateRatio} are
somewhat overestimated. Nevertheless, it is instructive to compare the
present ratios, shown in black, to those from previous work
\citep{ANG99,JAE01}, displayed in red. It can be seen that the present
ratio is significantly larger than previous results and, consequently,
we predict that more neutrons will be produced per captured
$\alpha$-particle.

\section{Astrophysical Implications}
\label{sec:results}

\subsection{Models}
\label{sec:num-method}
In order to explore how the current \Nepa reaction rates affect
s-process nucleosynthesis, two kind of calculations are presented
here. The first compares final abundance yields from post-processing
models upon changing the \textit{recommended} \Nepa reaction rates
from previously published results to those presented in this
paper. 
The second calculation estimates the variations in s-process
nucleosynthesis arising from \textit{uncertainties} in the present
\Nepa reaction rates. These can then be compared with abundance
variations arising from the literature rates.

In order to take the uncertainties into account, three sets of
calculations were performed: (i) recommended rates for both \an and
\ag reactions, (ii) low \an rate and high \ag rate, and (iii) high \an
rate and low \ag rate. Although in reality the \ag and \an rates will
be correlated, it is difficult to account for these correlations since
the \ag rate includes resonances below the neutron threshold and since
some \nuc{26}{Mg} levels contribute more to one reaction channel than
the other. For these reasons, we have chosen in the present study to
explore conservatively the impact of the largest reaction rate
variations. These nucleosynthesis calculations are performed
separately for massive stars and AGB stars.

\subsubsection{Massive Star Models}

A single zone temperature-density profile has been used to study the
effects of the \Nepa reaction rates on nucleosynthesis during the core
helium burning stage in massive stars. 
The temperature-density profile and initial abundances used in the
present study are for a $25M_{\odot}$ star and have been used
previously in Refs.\ \cite{THE00,ILIBook}.  The most abundant isotopes
at the onset of helium burning are (in mass fractions, $X$): $^4$He
($X_{\alpha}=0.982$), $^{14}$N ($X_{^{14}\text{N}}=0.0122$), $^{20}$Ne
($X_{^{20}\text{Ne}}=0.0016$), and $^{60}$Fe
($X_{^{60}\text{Fe}}=0.00117$). During most of the core helium burning
phase, the temperature and density ($T \approx 100 - 250$~MK and $\rho
\approx 1000 - 2000$~g$/$cm$^3$, respectively) are not high enough for the
\an neutron source to produce a significant number of
neutrons. However, towards the end of this phase the temperatures
become high enough for efficient neutron production. The exact time at
which the \an reaction starts to occur not only affects the number of
neutrons produced during core helium burning, but also the amount of
\nuc{22}{Ne} remaining that can be processed later during the carbon
shell burning phase. Although not studied here, the s-process is also
expected to be active during shell carbon burning.

The nucleosynthesis study was performed with a 583 nucleus s-process
network that extends up to molybdenum. Reaction rates (other than the
\Nepa rates) were adopted from the \texttt{starlib} library
\citep{ILI11}. The \texttt{starlib} library incorporates a compilation
of recently evaluated experimental Monte Carlo reaction rates in
tabular format on a grid of 60 temperatures from 1~MK to
10~GK. Tabulated are the temperature, the reaction rate, and the
factor uncertainty, which is closely related to the lognormal
parameter, $\sigma$, in Ref.\ \cite{LON10}.

\subsubsection{AGB Star Models}

The AGB nucleosynthesis tests are performed on a 5.5$M_{\odot}$,
$Z=0.0001$ model star, detailed in Ref.\ \cite{KAR10}. This model was
chosen because it experiences many thermal pulses (77 in total) during
the AGB phase, where 69 of those He-shell instabilities reach peak
temperatures of 0.30~GK or higher (with temperatures of 0.35~GK for 50
thermal pulses). 
One complication arises from disentangling the effects of the \Nepa
rates and those of proton-capture nucleosynthesis at the base of the
convective envelope (hot bottom burning, HBB). In our model, the base
of the envelope reaches peak temperatures of 98 MK, easily hot enough
for activation of the NeNe and MgAl proton-burning chains.  The main
results were reductions in the envelope \nuc{24}{Mg} and \nuc{25}{Mg}
abundances, and increases in \nuc{26}{Mg}, \nuc{26}{Al}, and
\nuc{27}{Al}.  This means that the He-intershell preceding a pulse
contains a non-solar Mg isotopic composition that is enriched in
\nuc{26}{Mg}.

The post-processing nucleosynthesis used for the AGB star models has
previously been described in detail by Ref.\
\cite[e.g.,][]{KAR10}. This code needs as input from the stellar
evolution code variables such as temperature, density, and convective
boundaries as a function of time and mass fraction. The code then
traces the abundance changes as a function of mass and time using a
nuclear network containing 172 species (from neutrons to sulphur, and
then from iron to molybdenum) and assuming time-dependent diffusive
mixing for all convective zones \citep{CAN93}. Although this network
does not contain species of the main s-process above A $\approx 100$,
their production is estimated by the inclusion of an extra isotope
(the ``g particle'', counting neutron captures beyond our
network). The reaction rates used in the nuclear network are mostly
taken from the JINA \texttt{reaclib} database \citep{CYB10}, with the
exception of the \Nepa rates adopted from the present work. Some
modifications were made to the JINA \texttt{reaclib} library including
the removal of the \nuc{96}{Zr} decay rate (since this is an
essentially stable isotope with a half-life of $t_{1/2} \ge 10^{19}$
years), and the inclusion of the ground and isomeric states in
\nuc{85}{Kr}. This is done because 50\% of the neutron flux from $n +$
\nuc{84}{Kr} proceeds to the ground state of \nuc{85}{Kr} ($t_{1/2} =
3934.4$ days) and the other 50\% goes to the isomeric state ($\tau =
4.480$ hours). The inclusion of both \nuc{85}{Kr} states is essential
for Rb abundance predictions in AGB nucleosynthesis models \citep[see
discussion in][]{GAR09,LUG11}.


\subsection{Results}
\begin{figure*}
  \centering
  \includegraphics[width=0.8\textwidth]{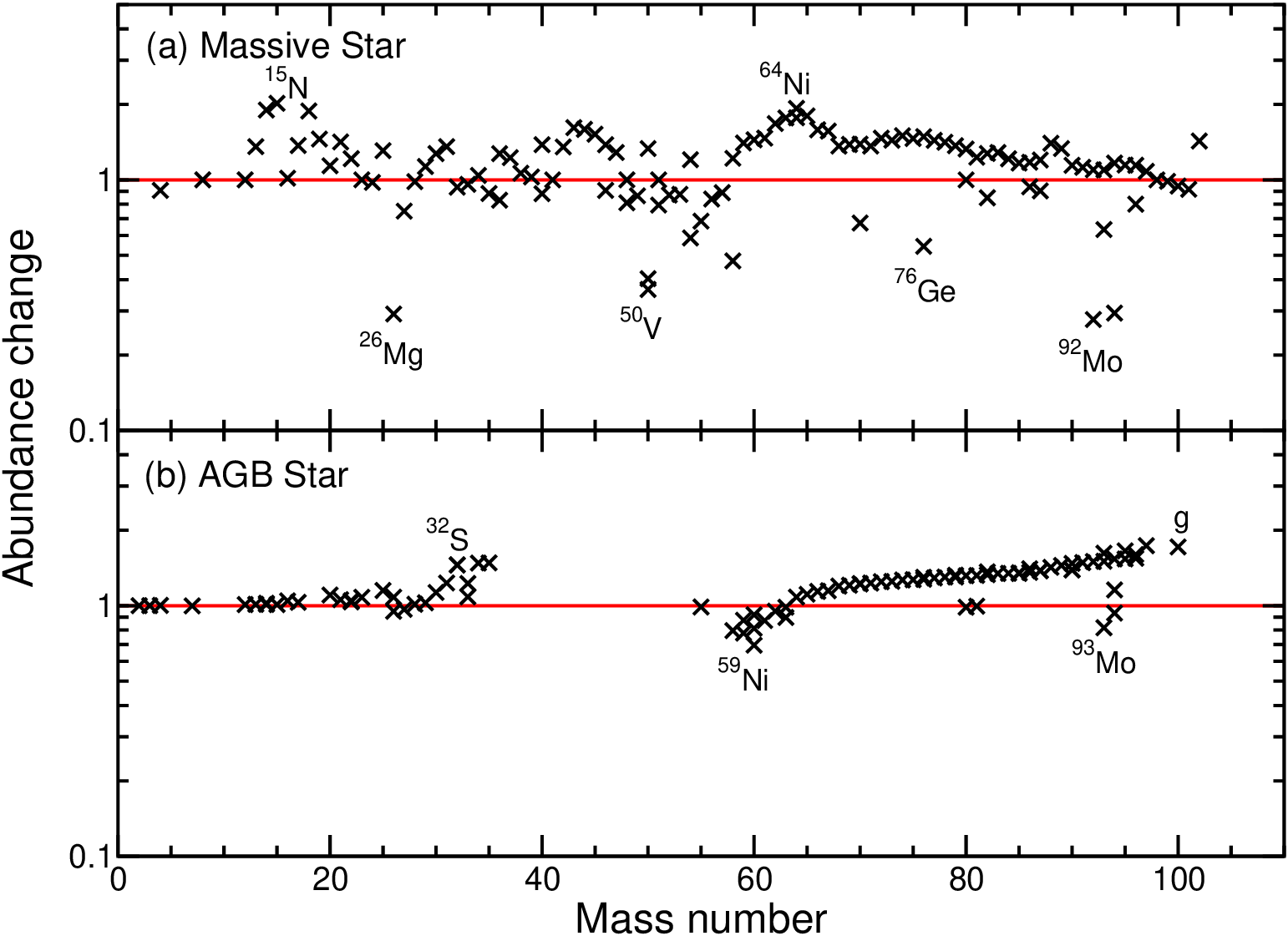}
  \caption{(Colour online) Ratio of final abundances resulting from
    the new recommended \ag and \an rates to those obtained from the
    old recommended rates. Points above unity (red line) represent a
    net increase in abundance. (a) At the end of core He-burning in a
    $25M_{\odot}$ star, the most significant abundances affected by
    the new rates are those of \nuc{26}{Mg} and the p-nuclei
    \nuc{74}{Se}, \nuc{78}{Kr}, and \nuc{84}{Sr}. (b) For AGB stars,
    higher mass nuclei are produced in larger quantities, as evidenced
    by the `g' particle that monitors neutron captures above
    molybdenum.}
  \label{fig:AbundanceChange-both}
\end{figure*}

\begin{figure*}
  \centering
  \includegraphics[width=0.8\textwidth]{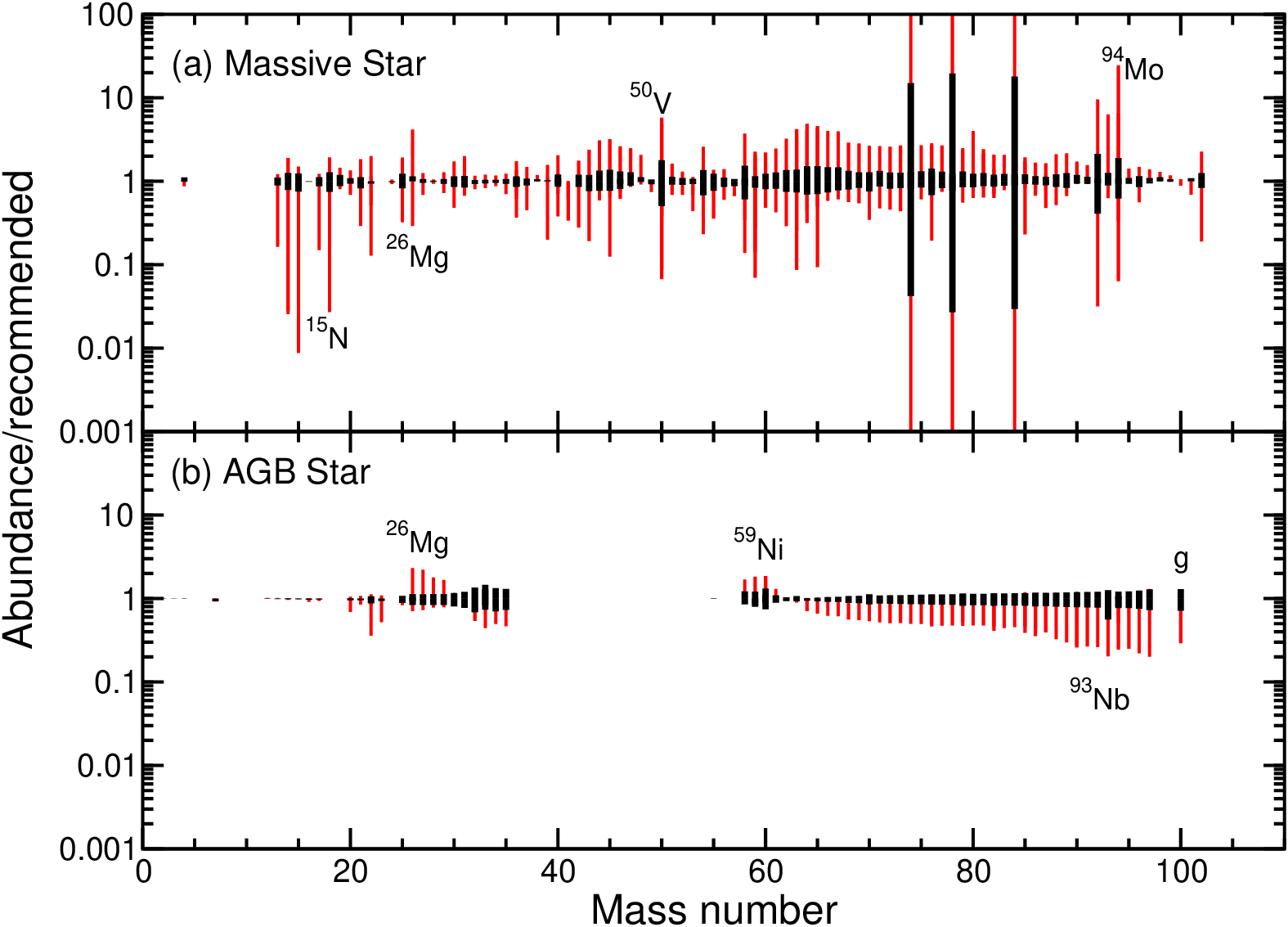}
  \caption{(Colour online) Comparison of abundance variations versus
    mass number arising from \Nepa rate uncertainties presented in the
    literature (Ref.\ \cite{ANG99} for the \ag reaction rate and Ref.\
    \cite{JAE01} for the \an reaction rate) and from the current \Nepa
    rate uncertainties. Abundance changes based on the present and
    previous rate uncertainties are shown as thick black bars and thin
    red bars, respectively, for (a) massive stars, and (b) AGB stars.}
  \label{fig:Uncertainties-both}
\end{figure*}

The effects of our new rates on the nucleosynthesis in comparison to
using the results obtained in the literature are shown in Fig.\
\ref{fig:AbundanceChange-both}. The improvements in abundance
predictions for the two stellar environments are shown in Fig.\
\ref{fig:Uncertainties-both}. The most up-to-date previously published
rates for the \ag and \an reactions are from Refs.\ \cite{ANG99} and
\cite{JAE01}, respectively. The effects are markedly different for the
two s-process environments, hence they will be discussed separately in
the following.

\subsubsection{Massive Stars}
The recommended \an reaction rate has not changed significantly in the
present analysis. Consequently, we do not expect the final
\nuc{25}{Mg} abundance to change. The final \nuc{26}{Mg} abundance, on
the other hand, changes significantly by roughly a factor of
three. The abundance changes in nuclei heavier than iron are smaller,
with the largest abundance increases occurring near \nuc{64}{Ni}. The
increased destruction of isotopes already present in the star is also
apparent for the p-nuclides \nuc{74}{Se}, \nuc{78}{Kr}, \nuc{84}{Sr},
and \nuc{93}{Nb}. These results indicate that with the reduced \ag
rate, more neutrons are produced per \Nepa reaction. Rather than
extending the reach of the weak s-process component (i.e., synthesis
of more massive nuclei), this flux increase affects branchings in the
s-process path close to the iron peak. A wider range of intermediate
mass nuclei are therefore produced. Fig.\
\ref{fig:AbundanceChange-both} also illustrates that the \Nepa rates
not only affect the abundances of traditional s-process nuclides, but
also the abundances of nuclei below the iron peak that act as
poisons. An example is \nuc{25}{Mg}, which produces \nuc{26}{Mg} through
the \reaction{25}{Mg}{n}{$\gamma$}{26}{Mg} reaction. With a higher flux
of available neutrons, this neutron poison reaction occurs more
frequently, effectively lessening the impact of the increased neutron
flux on s-process nucleosynthesis.

Uncertainties in s-process nucleosynthesis in massive stars arising
from uncertainties in the \Nepa reaction rates are shown in Fig.\
\ref{fig:Uncertainties-both}, where the thin (red) bars show
uncertainties arising from the old rates, and thicker (black) bars
show those from the new rates. In particular, large reductions are
noticeable for \nuc{26}{Mg}, where the current yield uncertainty
amounts to around 50\% in contrast to the previous factor of
5. Uncertainties in weak s-process nucleosynthesis have also undergone
significant improvements, especially for species that can only be
destroyed, but not created, by neutron captures. An example of this is
the nucleus \nuc{58}{Ni} whose yield uncertainty has been reduced from
a factor of five to just 50\%. It is important to note
here that, although the Monte-Carlo reaction rates do take into
account systematic uncertainties, it is difficult to account for
ambiguities in the data, for example, the open question of whether or
not the \Erlab{830} resonance is a doublet. Clearly, more measurements
are needed.

\subsubsection{AGB Stars}
Nucleosynthesis yields from our low metallicity AGB star models show a
very different pattern to those of the massive star study. For AGB
stars, the effect on lighter elements is reduced in comparison to
massive stars, with higher mass s-process elements revealing the
largest changes. This weighting toward higher mass s-process elements
is caused by our choice of using a low metallicity model. At low
metallicity, the neutron/Fe seed ratio is much higher meaning that
there is a higher production of higher atomic mass nuclei (e.g., see
discussion in Refs.~\cite{BUS01,KAR12}). Nuclei towards the upper end
of our network are produced up to a factor of 2 more than before, with
the `g' particle representing nuclei beyond our network capturing over
70\% more neutrons. In low metallicity AGB stars, therefore, the \Nepa
reactions can be expected to produce more high-mass s-process
elements, while leaving the low-mass s-process below $A \approx 80$
largely unaffected.

Uncertainties in s-process nucleosynthesis have been, as in massive
stars, dramatically improved with our new rates. The previous
abundance uncertainties were approximately a factor of 10, while the
present uncertainties amount to less than a factor of 2. The present
uncertainties in the rates affect the lower masses from A$\approx$25
to A$\approx$35 more than the s-process abundances. The ratio of
\nuc{26}{Mg} and \nuc{25}{Mg} is still uncertain by
approximately 20\%, whereas it was previously around 80\% (note that
Ref.\ \cite{KAR06} found \nuc{26}{Mg}/\nuc{25}{Mg} ratio uncertainties
of 60\%). Rubidium and zirconium isotopes have undergone yield
uncertainty improvements by a factor of about two. For the s-nuclide
\nuc{96}{Mo}, the uncertainty has been reduced from a factor of 4 to
a factor of 2 with our present results.

The new \Nepa reaction rates presented here should also be tested with
low-mass AGB star models ($M \le 3 M_{\odot}$). In lower mass AGB
stars, while the \reaction{13}{C}{$\alpha$}{n}{16}{O} reaction is the
main neutron source active between thermal pulses, activation of the
\Nepa reactions during a convective thermal pulse can have a
significant effect on branchings in the s-process path.

\section{Conclusions}
\label{sec:conclusions}

Both the \an and the \ag reactions influence the neutron flux
available to the s-process in massive stars and AGB
stars. Uncertainties in the rates, therefore, lead to large
uncertainties in s-process nucleosynthesis. In this paper, we have
estimated greatly improved \Nepa reaction rates, based on newly
available experimental information published since the works of Refs.\
\cite{JAE01} and \cite{ANG99}, and by applying a sophisticated rate
computational method\ \citep{LON10}. Subsequently, we explored the
astrophysical consequences for massive stars and for AGB stars.

In massive stars, simple one zone models of core helium-burning were
utilised to determine the influence of the new rates on the weak
component of the s-process. The most important result of our study is
a significant reduction of nucleosynthesis uncertainties. The yield
uncertainty has been reduced by between a factor of 5 and 10 across
the s-process mass region considered here (A $< 100$). For example,
the yields of key isotopes, \nuc{26}{Mg} and \nuc{70}{Zn}, have
uncertainty reduction factors of about 5 and 10, respectively.  When
comparing abundances obtained from our new recommended rates with
those derived from previous recommended rates, the final yield of
\nuc{26}{Mg} is found to have been reduced by roughly a factor of
three, while s-process isotopes were affected only
marginally. However, s-process nucleosynthesis is more concentrated
around the iron peak when using the new reaction rates. This relative
insensitivity to changes in neutron flux is partially caused by
captures on the neutron poisons \nuc{12}{C}, \nuc{16}{O}, and
\nuc{25}{Mg}, which are present in large quantities.

In our AGB star models, the final abundance uncertainties have also
been improved significantly with the new rates, with reductions by up
to an order of magnitude. The key rubidium and zirconium isotopes, for
example, have undergone yield uncertainty improvements of roughly a
factor of two. We have also found that s-process nucleosynthesis is
more active when including the new \Nepa reaction rates. While only
small changes are found in the low-mass s-process path ($A < 80$), at
higher masses production increases by up to a factor of 2. This is
especially evident by counting the number of captures at the end of
our network, yielding an increase of over 70\%. Further calculations
should be performed to study the effect of our new rates on lower mass
AGB stars, while paying special attention to their effects on
branchings in the s-process path.

The Monte-Carlo method used in the present study to calculate the
\Nepa reaction rates has the distinct advantage of calculating the
uncertainties in a robust and statistical meaningful manner. Although
our rates include some of the systematic uncertainties in the nuclear
data, there are still open questions regarding the resonance
properties that could affect the rates. Clearly, the remaining
ambiguities in the nuclear data for the \Nepa reaction rates need to
be resolved. The discrepancies discussed here, by \textcite{KOE02},
and by \textcite{KAR06}, make it difficult to assign some \nuc{26}{Mg}
levels to \Nepa resonances. Furthermore, the \Erlab{831} resonance
should be re-measured with high precision. More information should
also be gathered on the structure of \nuc{26}{Mg} levels near the
$\alpha$-particle and neutron thresholds. Indirect methods such as
particle transfer measurements are useful here, since the Coulomb
barrier inhibits direct measurements.

\section{Acknowledgements}

This work was supported in part by the US Department of Energy under
grant DE-FG02-97ER41041 and the National Science Foundation under
award number AST-1008355.  This work was also partially supported by
the Spanish grant AYA2010-15685 and the ESF EUROCORES Program
EuroGENESIS through the MICINN grant EUI2009-04167. AIK thanks Maria
Lugaro and Joelene Buntain for help with setting up the
nucleosynthesis code that reads in tables.  AIK is grateful for the
support of the NCI National Facility at the ANU.  RL would like to
thank James deBoer for the in-depth discussions about properties of
\nuc{26}{Mg} levels and the \Nepa reactions.

%



\end{document}